\pdfoutput=1
\RequirePackage{lineno}

\documentclass[graphicx,amssymb,amsmath,enumerate,11pt]{article}
\usepackage{epstopdf} 
\usepackage{fancyhdr}
\usepackage{subfigure}
\usepackage{url}
\usepackage{amssymb}
\usepackage{amsmath}
\usepackage{amsfonts}
\usepackage{array,hhline,dcolumn} 
\fancyheadoffset[]{0.1cm}
\usepackage[sort&compress,numbers]{natbib}
\def\al{\alpha}

\def\ga{\gamma}
\def\de{\delta}

\def\th{\theta}

\def\ka{\kappa}
\def\la{\lambda}

\def\si{\sigma}

\def\ch{\chi}
\def\ps{\psi}
\def\om{\omega}
\def\Ga{\Gamma}
\def\De{\Delta}

\def\gaf{{\gamma}_5}
\def\gam{{\gamma}_{\mu}}

\def\ganu{{\gamma}^{\nu}}

\def\nue{\nu_e}
\def\numu{\nu_\mu}

\def\nuebar{\bar\nu_e}
\def\numubar{\bar\nu_\mu}

\def\nh^#1{{\hat N}^{#1}}

\def\mF_#1{{\cal F}_{#1}}

\newcommand{\beq}{\begin{eqnarray}}
\newcommand{\eeq}{\end{eqnarray}}
\def\to{\rightarrow}

\def\no{\nonumber}
\def\fr#1#2{\frac{#1}{#2}}

\def\uMeV{\mbox{MeV}}

\def\ueVt{\mbox{eV}^2}

\def\ueVt{\mbox{eV}^2}

\def\piz{\pi^{\circ}}
\def\pip{\pi^{+}}

  \usepackage{epsfig}
  \usepackage{url}
  \usepackage{authblk}

\begin{document}      

%

\begin{titlepage}
\title{\bf Beyond Standard Model Searches in the MiniBooNE Experiment}
\author[1]{Teppei Katori}
\author[2]{Janet Conrad}
\affil[1]{Queen Mary University of London, London, E1 4NS, U.K.} 
\affil[2]{Massachusetts Institute of Technology, Cambridge, MA 02139, USA}

\maketitle

\begin{abstract}
The MiniBooNE Experiment has contributed substantially 
to beyond standard model searches in the neutrino sector. 
The experiment was originally designed to test 
the $\De m^2\sim1~\ueVt$ region of the sterile neutrino hypothesis by observing 
$\nue$ ($\nuebar$) charged current quasi-elastic signals from a $\numu$
($\numubar$) beam. 
MiniBooNE observed excesses of 
$\nue$ and $\nuebar$-candidate events in neutrino and anti-neutrino
mode, respectively. 
To date, these excesses have not been explained within the neutrino Standard Model ($\nu$SM), 
the Standard Model extended for three massive neutrinos. 
Confirmation is required by future experiments such as MicroBooNE.     
MiniBooNE also provided an opportunity 
for precision studies of Lorentz violation.  
The results set strict limits for the first time 
on several parameters of the Standard Model-Extension, 
the generic formalism for considering Lorentz violation.    
Most recently, an extension to MiniBooNE running,  with a beam tuned in
beam-dump mode, is being performed to search for dark sector particles.
This review describes these studies,  demonstrating that short
baseline neutrino experiments are rich environments in new physics searches.
\end{abstract}

\end{titlepage}

\section{Introduction}

Across the particle physics community,  the mysterious
periodic-table-like nature of the Standard Model (SM) is 
motivating searches for new particles, new forces and new properties
of the particles that are known.   The neutrino sector is proving a
rich environment for these searches.   Having already found one 
beyond-Standard-Model (BSM) effect--neutrino mass \cite{PDG}--a series 
of experiments are pursuing other potential signals.        
Unlike the case of three-neutrino oscillation measurements within $\nu$SM,  
many of these searches are pursued over short baselines,  from a few meters to approximately
a kilometer.   The Mini Booster Neutrino Experiment (MiniBooNE)  
at Fermi National Accelerator Laboratory (Fermilab)
is an excellent example,  having contributed substantially 
to BSM studies.

This review describes the MiniBooNE BSM program.  We begin by
describing the experiment.  This is followed by a discussion of the
MiniBooNE cross section studies, which have been essential input to
both the BSM searches within this experiment, and also to other
experiments, including T2K most recently \cite{T2K_osc}.    We then
describe three searches:  the sterile neutrino search which motivated
the experiment,    Lorentz violation searches which set the first limits
on five neutrino sector parameters,  and the search for dark sector particles 
which is now being pursued with a re-configured beam.

\section{MiniBooNE experiment}

MiniBooNE (running from 2002-2012) 
was originally designed to test the LSND signal~\cite{LSND_osc}. 
In the LSND experiment, low energy (0 to 53~MeV) 
muon anti-neutrinos were produced by pion decay-at-rest (DAR), 
and were detected by the liquid-scintillator-based LSND detector at 31~m from the target. 
The observed 3.8$\sigma$ excess of $\nuebar$ candidate events could be interpreted
as oscillations in the $\De m^2\sim 1~\ueVt$ region within a simple two
massive neutrino oscillation hypothesis,
where the oscillation probability is given by:
\begin{equation}
P(\bar \nu_\mu \rightarrow \bar \nu_e) = \sin^2 2\theta \sin^2(1.27 \Delta
m^2 L/E).
\end{equation}
Here, $\theta$ and $\Delta m^2$ are oscillation parameters to control the amplitude and 
the period, respectively (further discussed in Section \ref{sec:osc}), 
$L$ is the distance from neutrino production
to interaction in meters, and $E$ is the energy of the neutrino in
MeV.     

An experiment which maintains the same $L/E$ ratio should observe an
oscillation probability consistent with LSND if the simple two neutrino model is a
good approximation of the underlying effect.    However, by employing
an average $E$ which is an order of magnitude larger then LSND, the systematic errors
associated with production and decay are quite different.
If $L$ is increased accordingly, and no signal is observed, this
rules out the two neutrino oscillation hypothesis of the LSND result.

\begin{figure}[tb]
\begin{center}
\includegraphics[scale=0.5]{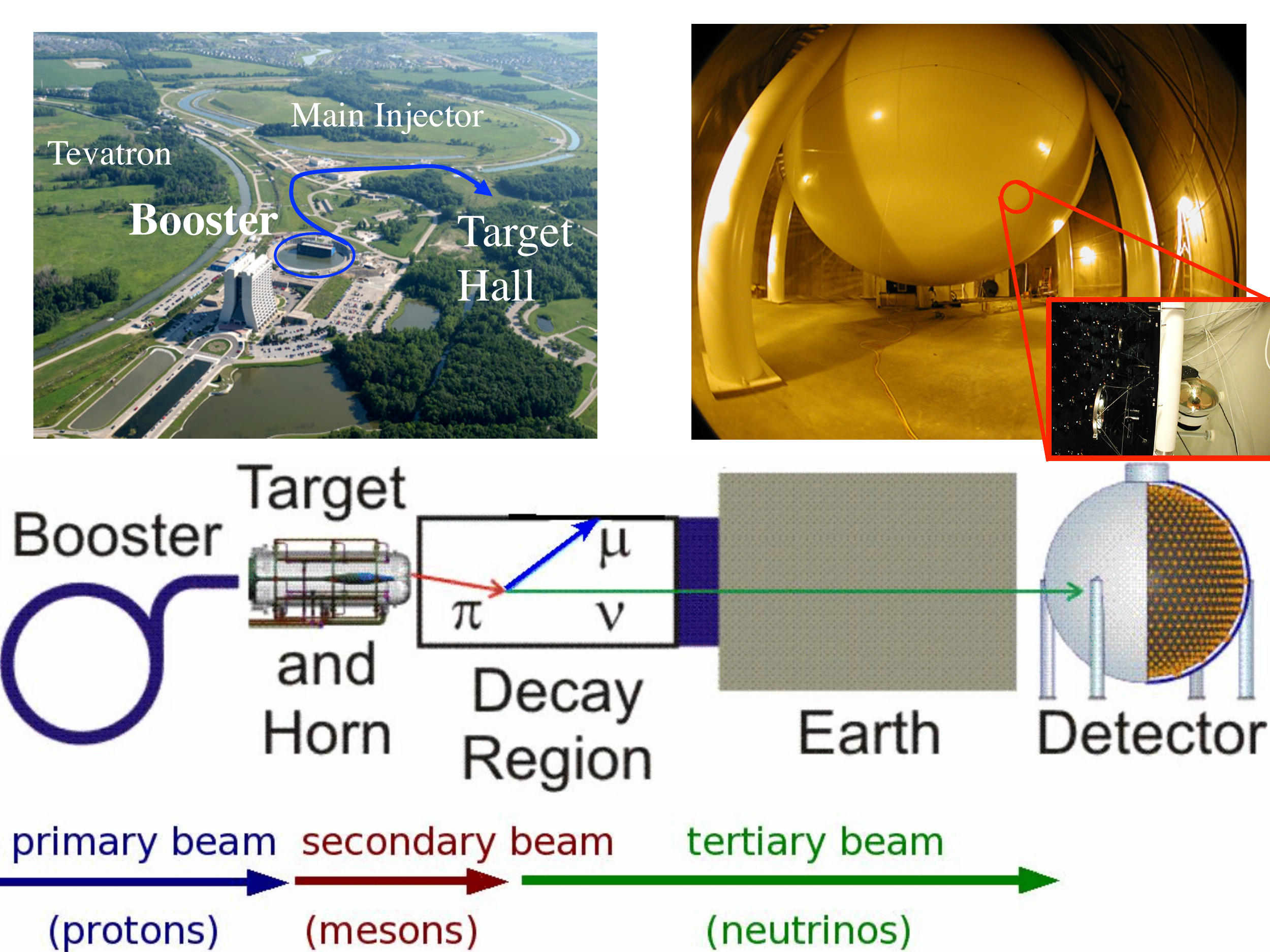}
\end{center}
\caption{\label{fig:MB_beam}
The MiniBooNE experiment layout~\cite{Teppei_LV}. 
Top left: the Fermilab accelerator complex.
Top right:  the MiniBooNE detector, with  
inset showing the black inner volume and the white outer volume.
Bottom: Schematic layout of the beam and detector~\cite{MB_CCQE}. 
 }
\end{figure}

MiniBooNE was designed with this in mind.  
The MiniBooNE beam peaked at $\sim$700~MeV and the Cherenkov detector was located at
$\sim$500~m baseline.
Fig.~\ref{fig:MB_beam} shows an overview of the MiniBooNE
design~\cite{Teppei_LV}, and in the remainder of this section we
provide more details.

\subsection{Booster Neutrino Beam-line~\label{sec:beam}}

The Booster Neutrino Beam-line (BNB) extracts 8~GeV kinetic energy protons from the Fermilab Booster, 
a 149~m diameter synchrotron (Fig.~\ref{fig:MB_beam}, top left).
Eighty one bunches,  separated in time by $\sim$19~ns,  are extracted
by a fast kicker within a $\sim 1.6~\mu$s pulse.  
Each pulse contains around $4\times 10^{12}$~protons.  
Typically, four to five pulses per second were sent to BNB to produce the neutrino beam. 

This high intensity proton pulse collides with
a beryllium target to produce a shower of mesons (Fig.~\ref{fig:MB_beam}, bottom). 
The target is located within a magnetic focusing horn.
For neutrino mode running, the toroidal field generated by the horn focuses positive mesons, 
with  $\pi^+$ decay-in-flight (DIF) as the primary source of the $\numu$ beam. 
In anti-neutrino mode running,
the horn focuses negative mesons to create the $\numubar$ dominant beam. 
The details of the BNB neutrino flux prediction can be found in
Ref.~\cite{MB_flux}.

MiniBooNE collected $6.46\times 10^{20}$~proton-on-target (POT) in neutrino mode, 
and $11.27\times 10^{20}$~POT in anti-neutrino mode.

\subsection{The MiniBooNE detector~\label{sec:detec}}

The MiniBooNE detector, located 541~m away from the target, 
is a mineral-oil-based Cherenkov detector. 
The 12.2~m spherical tank, filled with pure mineral oil, $(CH_2)_n$,
has two optically separated regions.  The interior region,
lined by 1280 8-inch photo-multiplier tubes (PMTs), contains the target volume. 
An outer volume, equipped with 240 8-inch PMTs, serves as
the veto region~\cite{MB_detec}. 
The presence of a charged particle above threshold is detected through the Cherenkov radiation observed by PMTs. 
As seen from Figure~\ref{fig:MB_beam}, top right, 
the inner volume is painted black to prevent scattering of the
Cherenkov light, 
improving the reconstruction precision.  On the other hand,
the outer volume is painted white to enhance scattering of Cherenkov light, 
in order to achieve the 99.9\% rejection of cosmic rays by the
veto~\cite{MB_firstosc} even with fairly sparse PMT coverage. 
The charge and time information from all PMTs is used to reconstruct kinematics of charged-lepton
and electromagnetic events. 
MiniBooNE mineral oil produces a small amount of scintillation light which can be used to reconstruct 
the total energy of the interaction via calorimetry, which is
particularly important for particles below Cherenkov threshold.

\begin{figure}[tb]
\begin{center}
\includegraphics[scale=0.5]{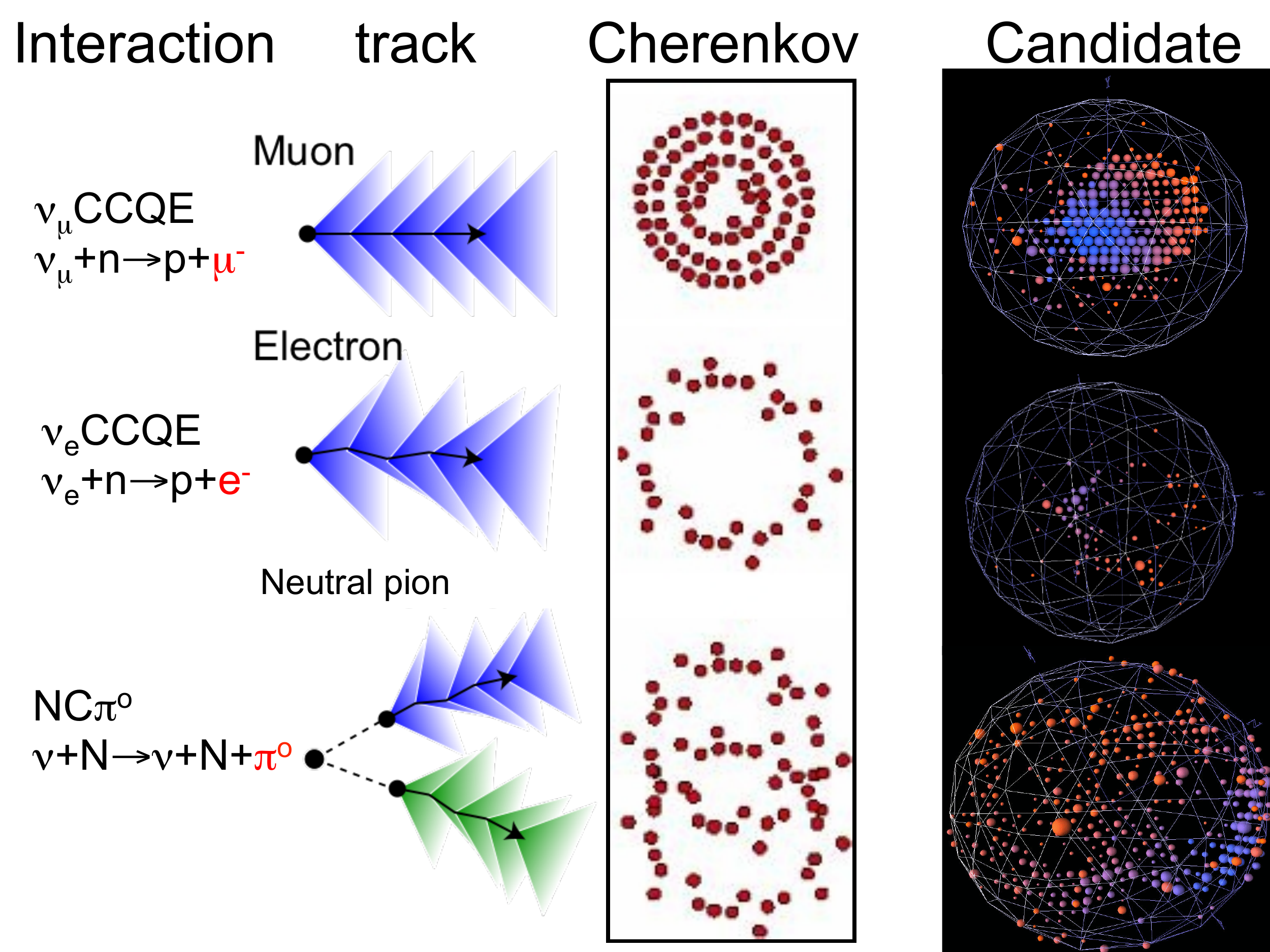}
\end{center}
\caption{\label{fig:MB_recon}
(color online) 
MiniBooNE particle reconstruction~\cite{Teppei_LV}. 
From top to bottom, a muon neutrino charged-current quasi-elastic (CCQE) interaction, 
an electron neutrino CCQE interaction, 
and a neutral current, neutral pion production (NC1$\piz$) interaction.  
The second and the third columns show the characteristics of tracks and Cherenkov rings~\cite{MB_firstosc}, 
and the last column shows the event displays of candidate events.
 }
\end{figure}

For the $\numu\to\nue$ ($\numubar\to\nuebar$) oscillation study, 
the following three particle reconstruction algorithms were the most important:  single
Cherenkov rings from 1) a muon and 2) an electron, 
and the two-ring electromagnetic shower topology from 3) a neutral pion decay to two gammas. 
Figure~\ref{fig:MB_recon} shows the different  characteristics of
these three signals, including examples of typical events in the detector
~\cite{Teppei_LV}. 
  
The reconstruction algorithms can also reconstruct more
complicated topologies important for constraining backgrounds and for cross section
studies  discussed below.
The charged-current single charged pion (CC1$\pip$) interaction reconstruction algorithm~\cite{MB_CCpip} 
fit two Cherenkov rings from final state particles, a charged lepton and a positive pion, to find their kinematics. 
The charged-current single neutral pion (CC1$\piz$) interaction reconstruction algorithm~\cite{MB_CCpi0} 
fit a charged lepton and a neutral pion (which consists of two electromagnetic showers, 
{\it i.e.} the algorithm fits for three Cherenkov rings).  
Another algorithm identifies and reconstructs the neutral current elastic (NCE) interaction~\cite{MB_NCE}, 
where the total kinetic energy of final state nucleons is found using scintillation light. 

Along with reconstruction of the light topology in the detector,
event identification also relies upon ``subevents.''  These are bursts of
light separated in time which indicate a sequence of decay.   For
example, a muon which stops and then emits a 
decay (``Michel'') electron will produce two subevents, one from the
initial muon and the one from the Michel electron.

\section{MiniBooNE cross-section results}

All searches for BSM physics rely on a precise understanding of 
SM interactions.    However,  when MiniBooNE began
running,  there was little neutrino cross section data in the 100 MeV to few 
GeV energy regime.
In response,  MiniBooNE developed a highly successful campaign of
cross section measurements, some of which are described here.  These results
are interesting by themselves and also can be used as direct inputs
to the BSM analyses, as described later in this paper.

MiniBooNE's beam is among the first high-statistics, high purity fluxes
in the energy range from 100 to 1500 MeV.        The observation of the resulting
events in a large, isotropic detector with $4\pi$ coverage is unique. 
Within this detector it is relatively easy to achieve uniform angular acceptance.
Also, the active veto makes it possible to measure NC interactions effectively.  
Insensitivity of hadronic details worked in positively. 
The hadron multiplicity often causes confusions for tracker detectors. 
Although the MiniBooNE detector cannot measure multiple hadron tracks, 
it measures total energy of low energy hadrons 
(such as protons below Cherenkov threshold from CCQE interactions) 
in calorimetric way, and, as a result, 
the details of final state interactions (FSIs), 
such as re-scattering, absorption, and charge exchange, 
do not strongly affect reconstruction of kinematics.

Perhaps most importantly to the overall impact of the data, 
the MiniBooNE collaboration provided the cross section data in a form 
that is most useful to theorists.
Traditionally, cross section data have been presented either as a function of 
neutrino energy ($E_\nu$) or 4-momentum transfer ($Q^2$). 
This presentation is problematic in the MiniBooNE 
energy region, because of the importance of nuclear effects:
Fermi motion smears the kinematics, binding energy shifts the energy spectrum, 
nucleon correlations affect both energy dependence and normalization of cross sections, 
pions may be created, absorbed, and charge-exchanged within the
nuclear  environment.
These nuclear processes modify the features of primary neutrino-nucleon interactions, 
and so model dependent corrections are required to reconstruct $E_\nu$ and $Q^2$. 
This model dependence is problematic because there are a wide range of
models available~\cite{NUANCE,NuWro,GENIE,NEUT,GiBUU}.

Instead, MiniBooNE chose to publish flux-integrated differential
cross sections in terms of measured kinematic variables,
which are essentially model-independent.
These results have the detector efficiency unfolded but are presented without any others corrections.  
In particular, the neutrino flux is not unfolded. 
The result is data that is neutrino-beam specific, and theoretical models are comparable 
only if those models  are convoluted with the MiniBooNE predicted neutrino flux. 
However, this is trivial for all theorists to do, 
given that MiniBooNE published a first-principles flux prediction~\cite{Sobczyk_dd}.  
This isolates all model dependence in the 
data-to-prediction comparison entirely to the ``prediction'' side of
the discussion.   The data remains completely general.
For this reason, the MiniBooNE cross section data are widely used to
study and compare theoretical models. 
In this section, we describe each cross section measurement briefly. 

\subsection{Charged-Current Quasi-elastic (CCQE) scattering~\label{sec:ccqe}}

\begin{figure}[tb]
\begin{center}
\includegraphics[scale=0.35]{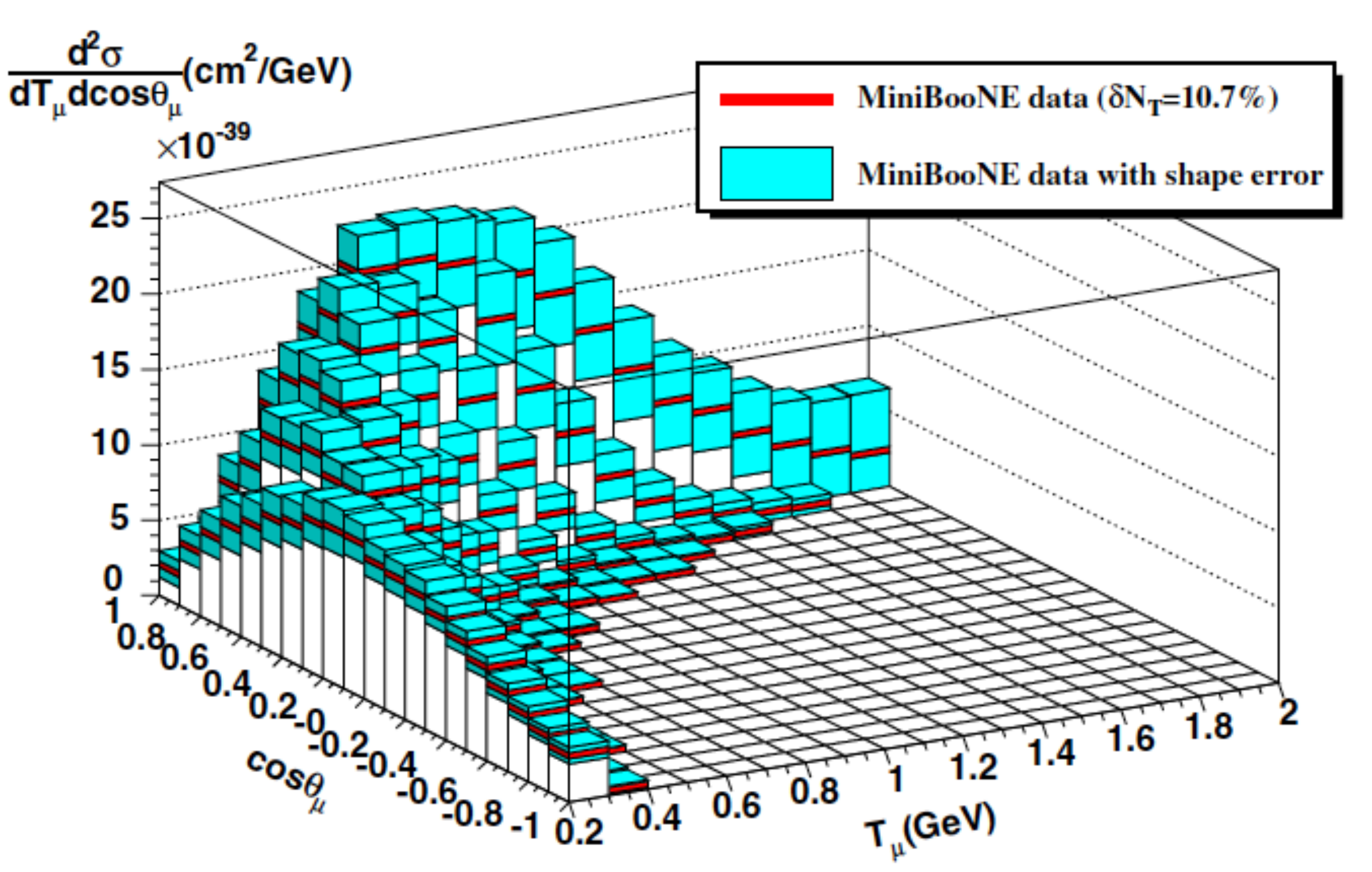}
\includegraphics[scale=0.35]{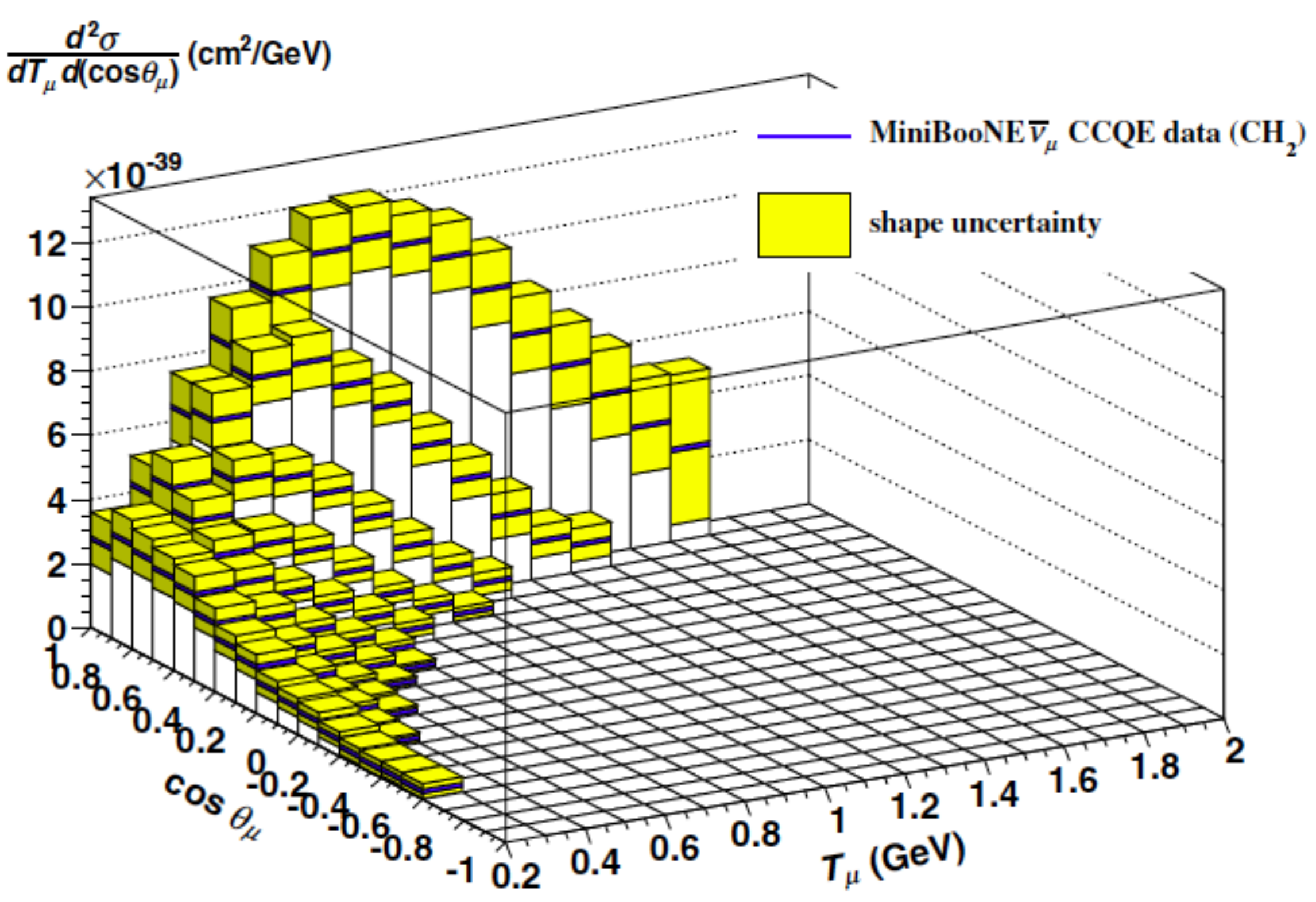}
\end{center}
\caption{\label{fig:ccqe}
(color online) 
MiniBooNE CCQE cross sections. 
The left plot shows the muon neutrino flux-integrated CCQE double differential cross section on a neutron target. 
The right plot shows muon anti-neutrino flux integrated CCQE double differential cross section 
on a CH$_2$ molecule. 
}
\end{figure}

The CCQE interaction is the primary interaction at MiniBooNE energies.
This interaction is used to detect $\numu$ ($\numubar$) and $\nue$($\nuebar$) candidate events 
in the oscillation and Lorentz violation analyses, 
\beq
\numu+n\to \mu^-+p~&,&~\numubar+p\to \mu^++n~,\no\\
\nue +n\to  e^- +p~&,&~\nuebar +p\to  e^+ +n~.\no
\eeq
Therefore, a strong understanding of this channel is essential.
High statistics $\numu$ ($\numubar$) interactions 
are used to study outgoing lepton kinematics~\cite{MB_CCQEPRL}. 
The observable of this channel is the outgoing muon, with no pions in the final state, {\it i.e.}, 
the signal event topology is ``1 muon + 0 pion + N protons''.
The main results were published in terms of  
flux-integrated double differential cross sections,
as functions of the lepton kinetic energy and the scattering angle. 
Figure~\ref{fig:ccqe} left shows the flux-integrated double differential cross section 
of $\numu$CCQE interactions~\cite{MB_CCQE}. 
The irreducible background from the pion production channel 
is subtracted based on a sideband study, 
but the subtracted background is also published 
so that readers can recover the irreducible background.

These data have revealed the importance of nucleon correlations~\cite{Martini_first,Nieves_first} 
in neutrino scattering,
which had not been taken into account correctly in previous calculations. 
This led to models developed using electron scattering data, that were
tested against MiniBooNE data~\cite{Martini_dd,Nieves_dd,Donnelly_dd,Bodek_tem,Meucci_first,Mosel_mec}. 
These models await being tested further by other experiments, 
such as MINERvA~\cite{MINERvA_nu,MINERvA_antinu} and T2K~\cite{T2K_CCincl}.  

Another important test is CCQE anti-neutrino scattering, where a wide
range of 
expectations 
were predicted prior to the run~\cite{Martini_antinu,Martini_antinudd,Nieves_antinu,Meucci_antinu,Donnelly_antinu}. 
Before the data could be compared to the results,
however, the substantial contamination of neutrinos in the
anti-neutrino beam had to be addressed.    
Three independent methods were used to constrain and
tune the neutrino contamination prediction~\cite{MB_ANTI}.
After subtracting the neutrino contamination, 
the flux-integrated double differential cross section 
for the muon anti-neutrino CCQE interaction was measured 
(Fig.~\ref{fig:ccqe}, right)~\cite{MB_ANTICCQE}. 
The comparison of models with data showed a preference 
for the high cross section models~\cite{Grange_NuInt12}.
The rich shape information of the double differential data continues
to provide additional tests,  beyond the normalization.

The main result of the $\numubar$ CCQE cross section measurements is quoted as per CH$_2$ molecule.    
This is because the MiniBooNE target consists of CH$_2$, and 
the experiment cannot distinguish 
anti-neutrino interactions with bound protons in the carbon nuclei 
and free protons from hydrogen.  
As a separate study, however,  MiniBooNE
also presented an analysis that subtracted the hydrogen interactions,
where the cross sections were then expressed per bound proton.   
This has also provided a useful handle for theorists.

\subsection{Charged single pion production~\label{sec:pion}}

The understanding of charged-current single-pion channels is of great interest to the nuclear community, 
but also, there are significant implications for the neutrino oscillation studies.
These interactions produce an 
irreducible background for CCQE events~\cite{Martini_erec,Martini_osc,Nieves_erec,Mosel_erec}.
If the detector fails to tag outgoing pions, either because of
detector effects or nuclear effects, 
pion production channels 
may be misclassified as CCQE. 
The distributions of irreducible backgrounds must be modelled,   
and those models rely on the pion production measurements, especially
the MiniBooNE data described here. 
Therefore, understanding the kinematic distributions of 
pion production channels is a crucial task for neutrino oscillation physics.

There are three pion production channels for which MiniBooNE performed dedicated measurements:
charged-current single $\pip$ (CC1$\pip$) production~\cite{MB_CCpip};
charged-current single $\piz$ (CC1$\piz$) production~\cite{MB_CCpi0}; 
and neutral current single $\piz$ (NC1$\piz$) production~\cite{MB_NCpi0}. 
\beq
\numu+CH_2&\to&\mu^-+\pi^++X~,\no\\
\numu+CH_2&\to&\mu^-+\piz+X~,\no\\
\numu(\numubar)+CH_2+&\to&\numu(\numubar)+\piz+X~.\no
\eeq
Here, the topologies of each event are more complicated and are described
as 
``1 muon + 1 positive pion + N protons'' (CC1$\pip$), 
``1 muon + 1 neutral pion + N protons'' (CC1$\piz$), and 
``0 muon + 1 neutral pion + N protons'' (NC1$\piz$). 
Although the MiniBooNE detector is not magnetized and 
therefore cannot distinguish positive and negative pions based on their
trajectories,  separation is possible. 
Negative pions are absorbed by a nucleus almost 100\% of the time, and in consequence, 
there is no emission of a Michel electron. 
This fact allows MiniBooNE to use the presence of a Michel electron to select positive pions. 

\begin{figure}[tb]
\begin{center}
\includegraphics[scale=0.35]{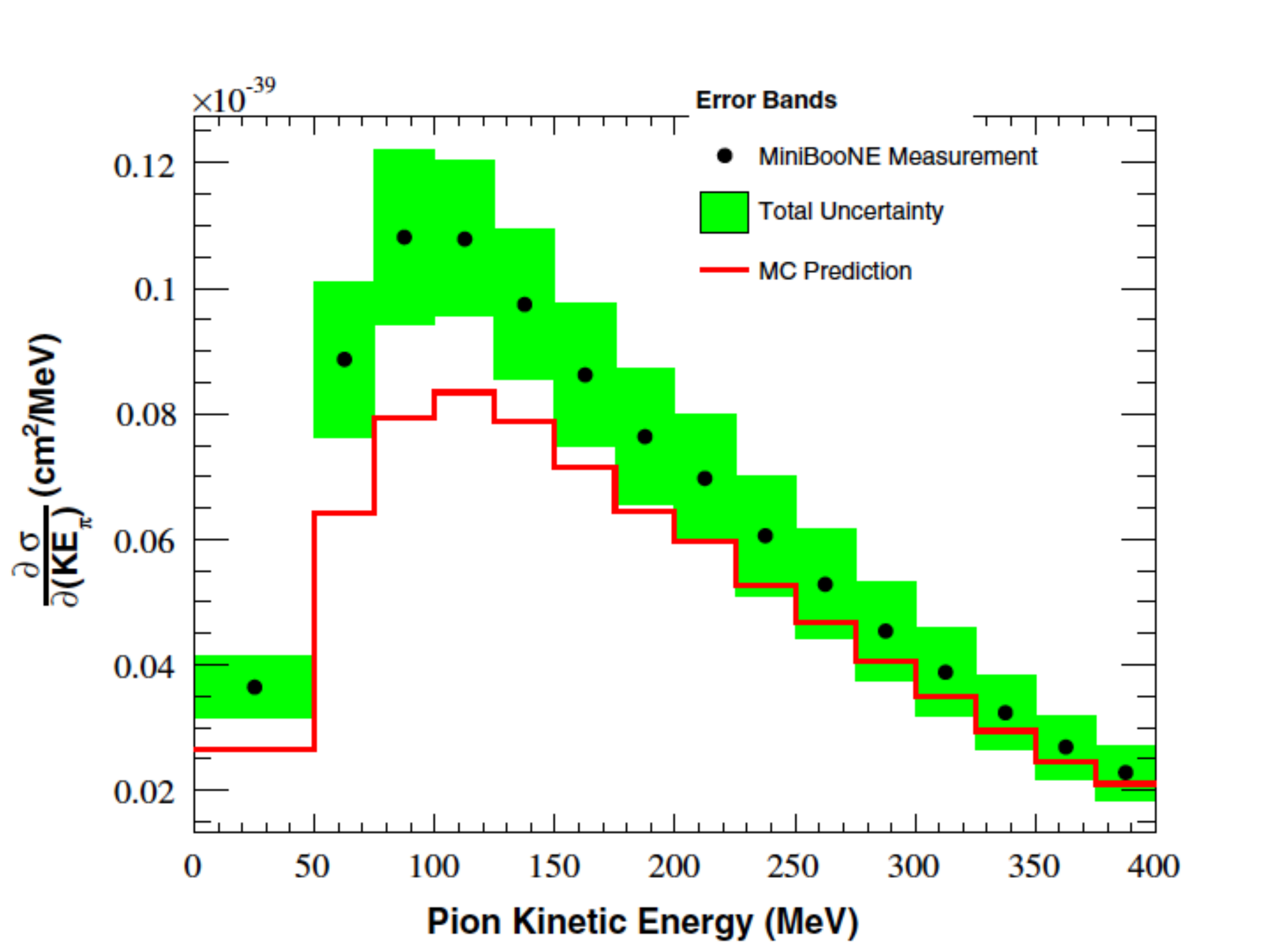}
\includegraphics[scale=0.35]{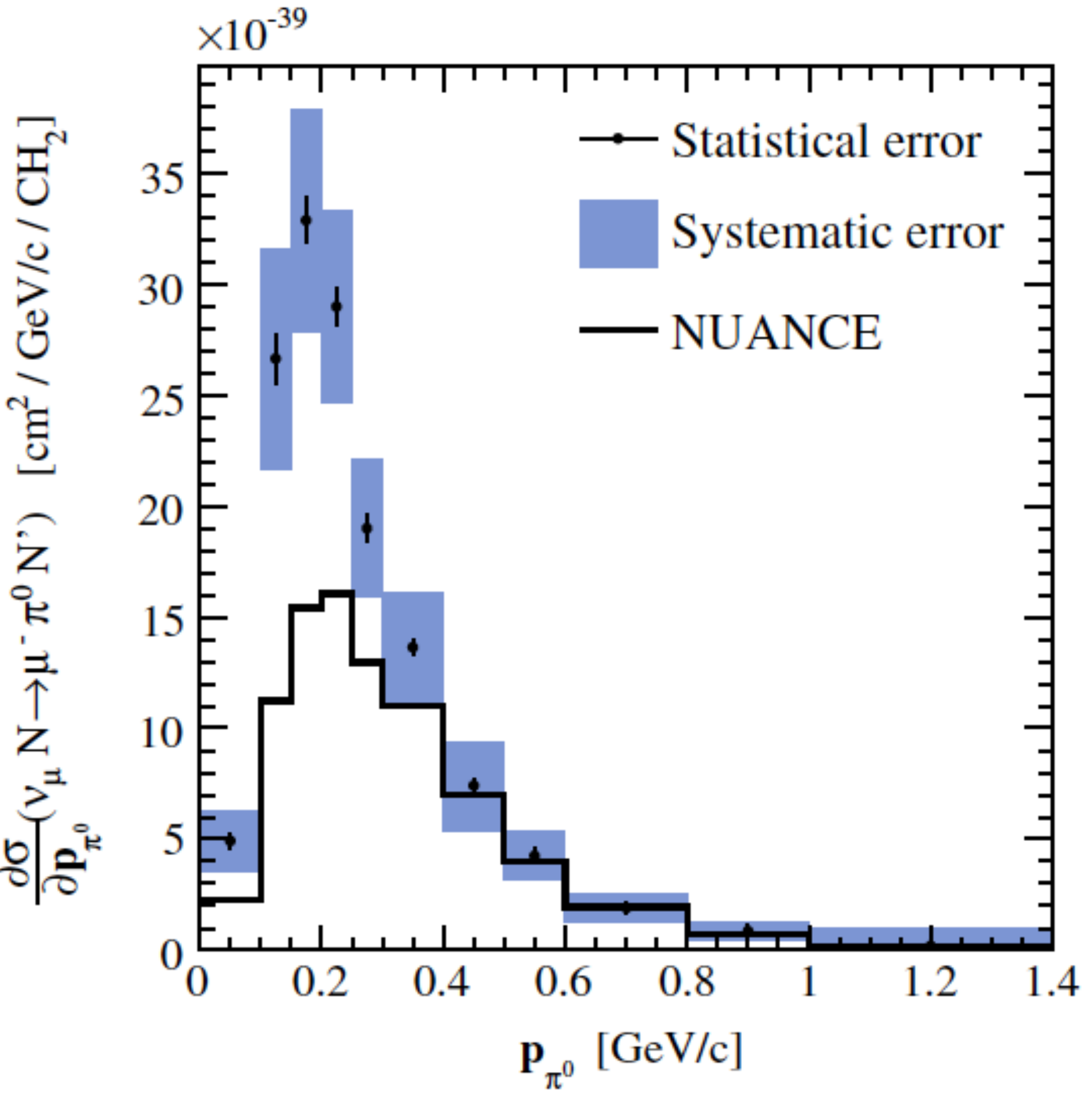}
\end{center}
\caption{\label{fig:pion}
(color online) 
MiniBooNE single pion production results. 
The left plot is $\pi^+$ kinetic energy differential cross section 
from CC1$\pip$ interaction on CH$_2$~\cite{MB_CCpip}. 
The right plot is $\piz$ momentum differential cross section 
from CC1$\piz$ interaction in CH$_2$~\cite{MB_CCpi0}. 
As you see, predictions underestimate data for both channels and the shapes do not agree as well. 
}
\end{figure}

Because of the more complicated topologies, 
the differential cross sections for these data sets are presented in various variables. 
Among them, distributions in pion kinetic energy and momentum distributions exhibit
the presence of nuclear effects, while 
we do not see this from the lepton distributions. 
Figure~\ref{fig:pion} shows differential cross sections, CC1$\pip$ pion kinetic energy 
and CC1$\piz$ pion momentum, respectively.  
The shape and normalization are sensitive to nuclear effects, such as pion absorption, charge exchange, 
and rescattering. 
Therefore, the state-of-the-art nuclear models~\cite{Mosel_pion,Hernandez_pion} can be tested by 
these MiniBooNE data.

\subsection{Neutral-Current Elastic (NCE) scattering~\label{sec:nce}}

\begin{figure}[tb]
\begin{center}
\includegraphics[scale=0.35]{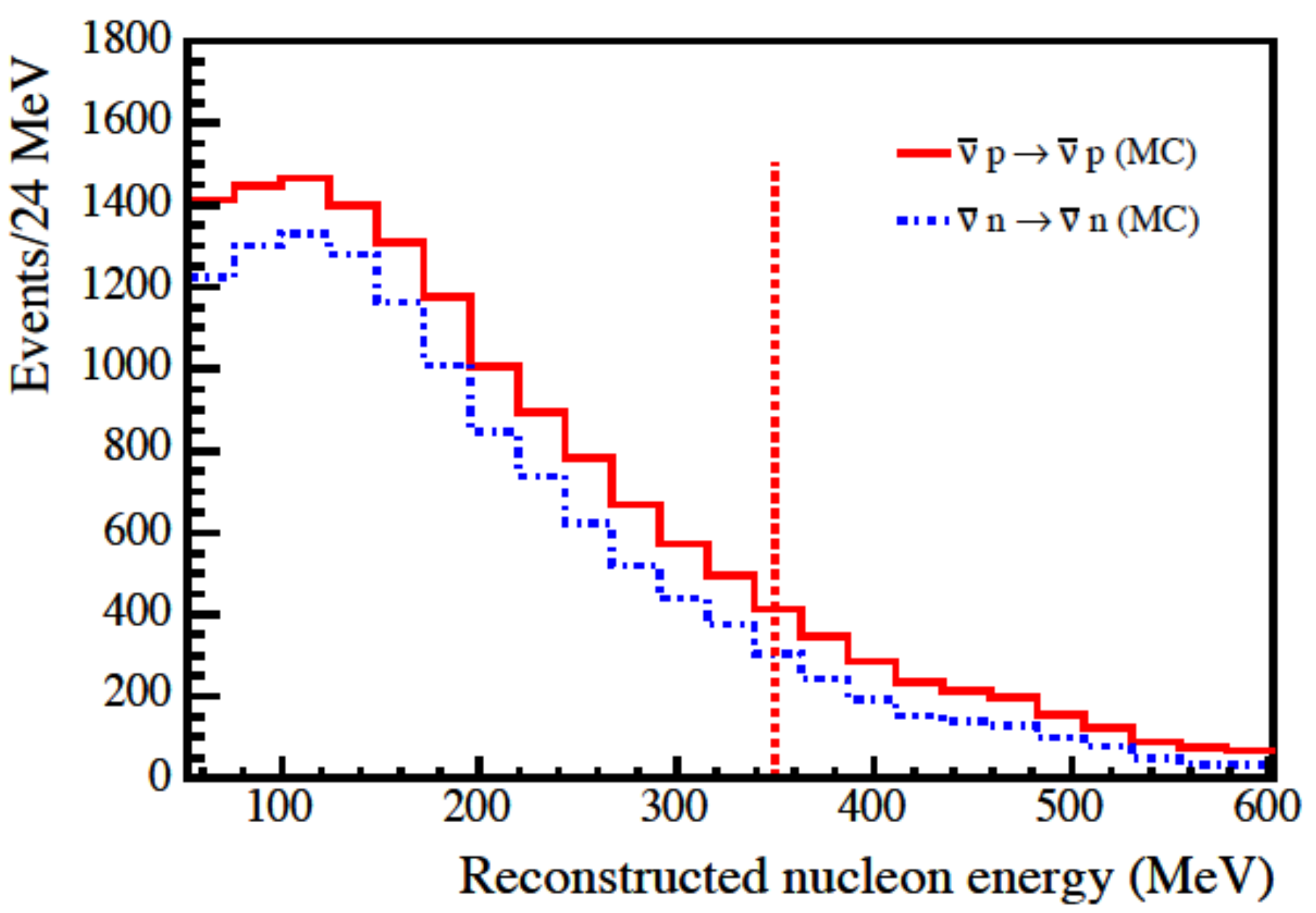}
\includegraphics[scale=0.35]{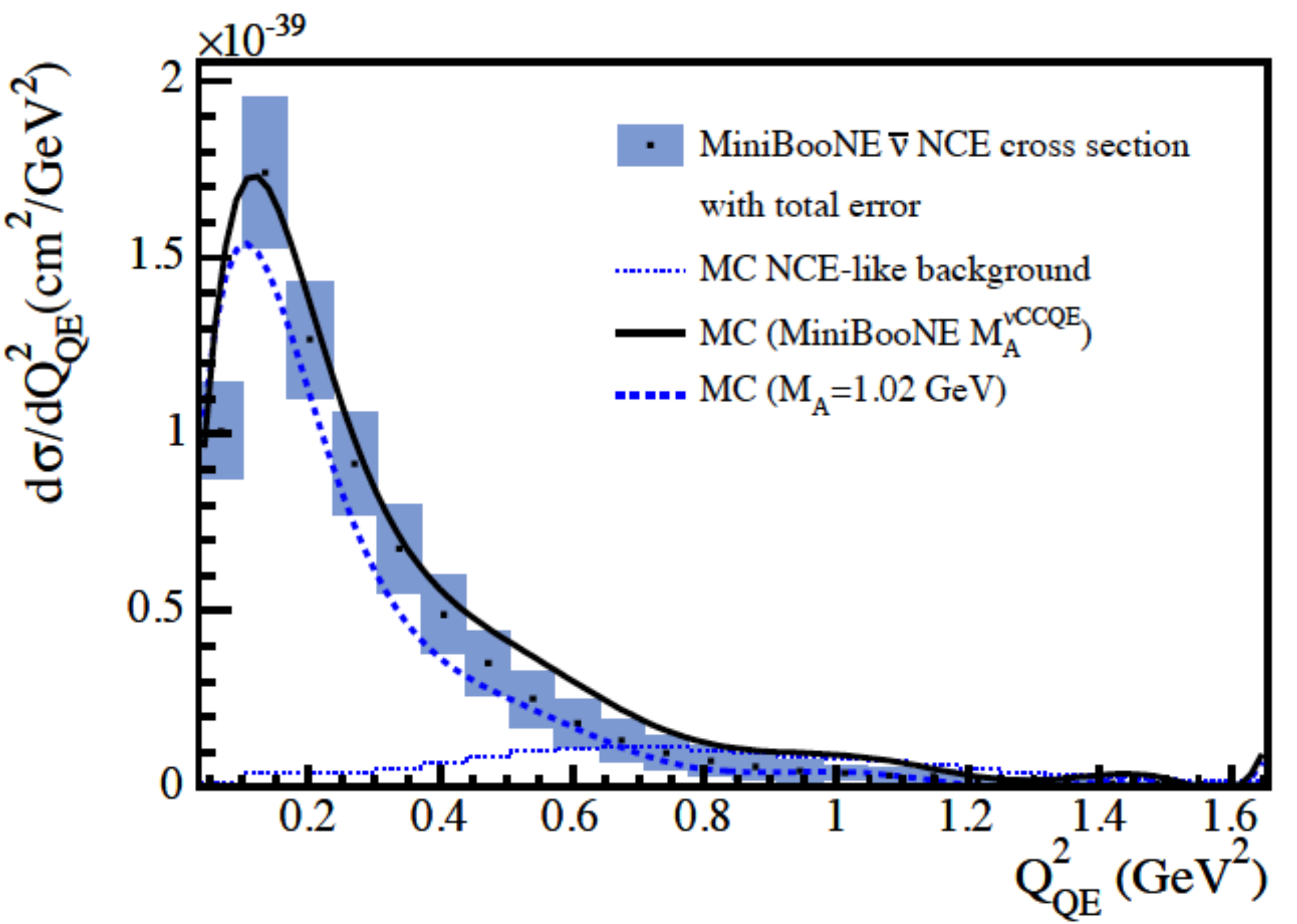}
\end{center}
\caption{\label{fig:nce}
(color online) 
MiniBooNE NCE results~\cite{MB_ANTINCE}. 
The left plot shows simulated kinetic energy of protons and neutrons from NCE in MiniBooNE. 
The line denotes the Cherenkov threshold, {\it i.e.}, 
only protons which have higher energy from this line emit Cherenkov radiation. 
For neutrons, there is no Cherenkov radiation and the chance the secondary proton from the primary neutron 
exceeds this threshold is extremely low 
(in other words, if the proton exceeds Cherenkov threshold, 
this will most likely form the primary neutrino NC interaction). 
The right plot shows the anti-neutrino NCE differential cross section. 
As you see, the data shows a ``roll-over'' in the low $Q^2$ region. 
}
\end{figure}

The NCE interaction can take place on both neutrons and protons,
for both neutrino and anti-neutrinos.  
The results are relevant for dark matter searches in two ways:
first through the measurement of $\De s$ that we describe here; 
second as a background to a direct dark matter search by MiniBooNE, 
described in sec.~\ref{DMsearch}. 
\beq
\numu(\numubar)+p&\to&\numu(\numubar)+p~,\no\\
\numu(\numubar)+n&\to&\numu(\numubar)+n~.\no
\eeq
Since only protons with kinetic energy above $\sim$350~MeV produce Cherenkov radiation 
(Fig.~\ref{fig:nce}, left), 
the  majority of these events only produce scintillation
light, and therefore necessitate a strictly calorimetric analysis. 
We call this topology ``0 muon + 0 pion + N protons''. 
However, 
when the kinetic energy exceeded the Cherenkov threshold, 
it is also possible to observe the direction of nucleons~\cite{MB_NCE}.  

The calorimetric measurement causes the signal to be 
insensitive to the detailed final state interaction (FSI) process. 
Also, similar to the anti-neutrino CCQE analysis (Sec.~\ref{sec:ccqe}), 
scattering on C and H cannot be distinguished, so the target may be 
a bound proton, a free proton, or a bound neutron.  Hence,
the cross section is presented per CH$_2$ target.
Figure~\ref{fig:nce}, right, shows the anti-neutrino mode 
NCE differential cross section~\cite{MB_ANTINCE}. 

The NCE data allows us to refine our understanding of nuclear effects at low $Q^2$. 
In NCE, the observable is the sum of all kinetic energies of out going protons, $\sum T_N$. 
Using this, the $Q^2$ can be reconstructed by assuming the target nucleon at rest,
\beq
Q^2_{QE}=2M_N\sum T_N 
\eeq  
Note that irreducible backgrounds, 
such as NC pion production without an outgoing pion, 
are subtracted to make $Q^2_{QE}$ physical.  

The reconstructed data shows a roll-over at the low $Q^2$ region, 
due to the combination of Pauli blocking and the nuclear shadowing. 
Pauli blocking is a phenomena where low momentum transfer interactions 
are forbidden due to occupied phase space, 
and the nuclear shadowing happens when 
the resolution (=low momentum transfer interaction) 
is insufficient to resolve a single nucleon wave function. 
Note that these nuclear effects do not appear if the signal of NCE is
defined to be a  single isolated proton, 
where strong FSI migrates all nucleons to low energy region~\cite{Mosel_nce}. 
However, because the MiniBooNE NCE data presents the sum of the total
nucleon kinetic energy, 
the results preserve the feature of the primary neutrino interaction physics.

NCE interactions are connected to direct dark matter
searches through the measurement of  $\De s$ , the spin of the strange
quarks in the nucleon. It has been shown~\cite{Ellis_Deltas} that the uncertainty of $\De s$ on 
the spin-dependent scattering between dark matter particles and target nuclei 
can be a large systematic error. Therefore, a $\De s$ measurement is another way that 
neutrino cross section measurements contribute to BSM physics.  We
briefly consider how this information can extracted from the NCE data here.

The spin structure of a nucleon is deeply fundamental and quite complicated.
In the naive constituent quark model, the
spin-$\frac{1}{2}$ of a nucleon can be derived by adding valence quark spins, 
where in the static limit ($Q^2\to 0$) 
there are three valence quarks that make up all static properties of a nucleon, 
such as charge, magnetic moment, and spin.   
However, the spin contribution from up and down quarks 
deduced from inclusive deep inelastic scattering (DIS) measurements~\cite{EMC,SMC,Vassili_ERC} 
indicate, in the static limit, that up and down quarks support only $\sim$10\% of the total spin of a proton. 
This so-called ``spin crisis'' has triggered a world wide effort to look for
other sources of spin in a nucleon. 
One of the interesting additional spin contributions is from the strange quarks, called $\De s$. 
Although recent measurements show the static limits of the strange quark charge 
and magnetic contributions are consistent with zero~\cite{G0}, 
the nonzero value of $\De s$ is still under debate~\cite{Pate}, 
because the weak coupling ($\propto (1-4sin^2\th_w)$) of $\De s$ with parity violating electron asymmetry 
does not allow a clear measurement of $\De s$ through electron scattering experiments.  

\begin{figure}[tb]
\begin{center}
\includegraphics[scale=0.4]{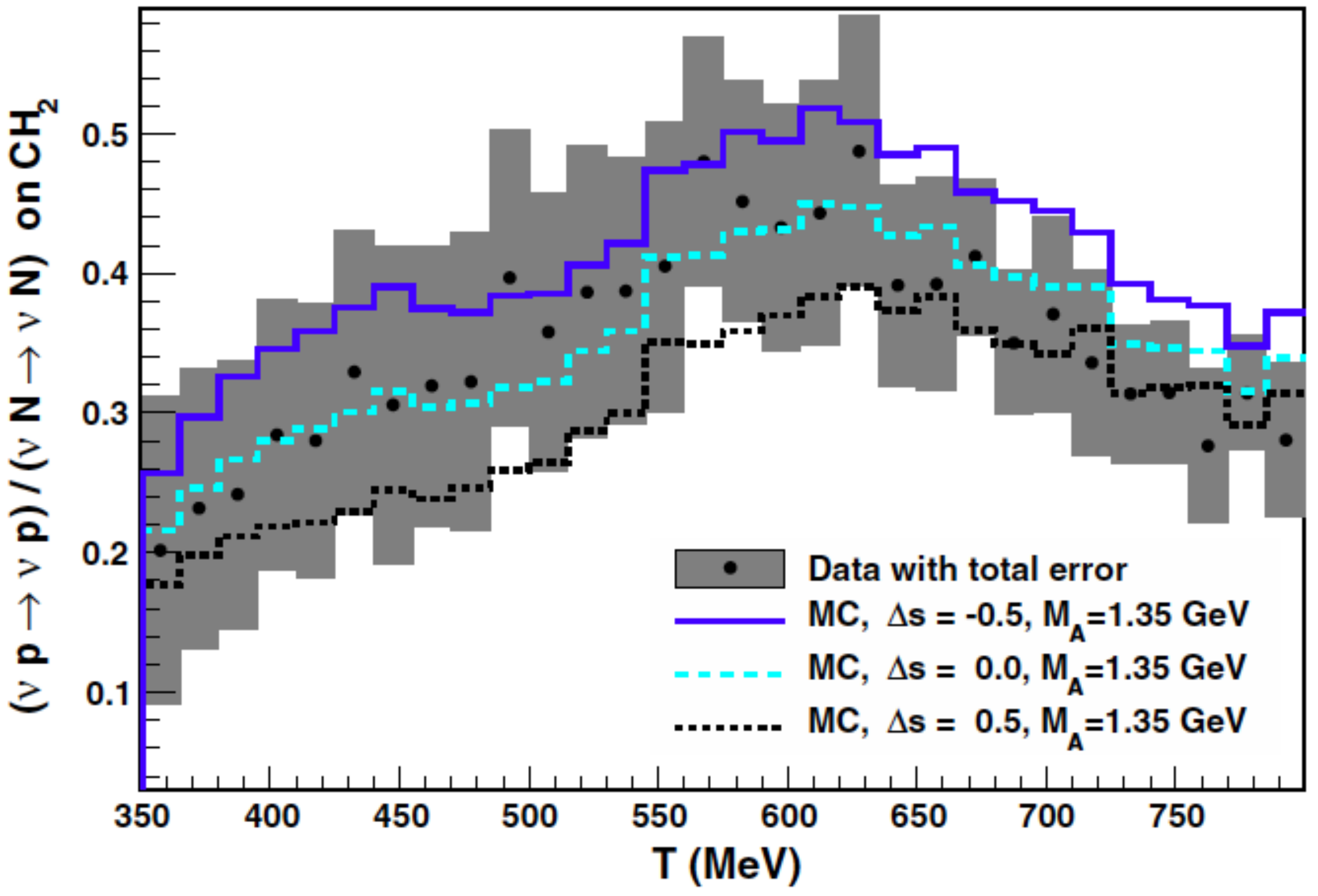}
\end{center}
\caption{\label{fig:deltas}
(color online) 
The MiniBooNE $\De s$ fit result~\cite{MB_NCE}. 
The contribution of $\De s$ increases the proton NCE cross section, and decreases the neutron NCE cross section. 
Therefore, it is necessary to separate proton and neutron NCE, 
which is possible for the high energy sample. 
Here, the denominator is chosen to be the total NCE events ($\nu+N\to\nu+N$) in order to cancel systematics. 
However, the large systematic errors do not allow MiniBooNE to 
definitively measure a nonzero $\De s$ contribution.
}
\end{figure}

However, $\De s$ also contributes to neutrino NCE scattering, as an axial vector isoscalar term,
increasing the cross section for neutrino-proton NCE, 
and decreasing the cross section for neutrino-neutron NCE.
MiniBooNE can only isolate neutrino-proton NCE in the case of high energy protons. 
Fig.~\ref{fig:deltas} shows the fit result of neutrino NCE data. 
The best fit value is $\De s=0.08\pm0.26$. 
Unfortunately, MiniBooNE does not have enough sensitivity to definitively determine nonzero $\De s$. 
This is due to the poor experimental proton-neutron separation 
which is only possible at high energy with large systematics. 
Therefore, a detector which has the ability to identify low energy protons, 
such as MicroBooNE~\cite{uB}, 
will have better sensitivity to $\De s$. 

\section{MiniBooNE oscillation results~\label{sec:osc}}

The most well-known BSM search performed by the MiniBooNE experiment
was for neutrino oscillations consistent with LSND.    These are also
the most thoroughly reviewed results.   Here,  we briefly describe the
studies.  We recommend Ref.~\cite{annualreview} for
a more extended discussion.

MiniBooNE was conceived in 1998,  shortly after the LSND results 
had reached 3.8$\sigma$ significance and
before the three massive neutrino model 
for active-flavor oscillations ($\nu$SM) had been well-established.  
However,  it was clear that if LSND was observing an oscillation
signal, the associated squared mass splitting ($\De m^2_{large}$) was more than an order
of magnitude larger than other evidence for oscillations.   In this
circumstance, a complicated three-neutrino
appearance probability can reduce to a more simple two neutrino 
case for designs with $(1.27 L/E) \approx 1/\De m^2_{large}$,
such as MiniBooNE.    

This approach assumes no $CP$ violation in the mixing matrix, 
and hence equal probabilities of neutrino and anti-neutrino oscillations. 
Leptonic $CP$ violation in the mixing matrix had been discussed 
by Wolfenstein in 1978~\cite{Wolfenstein:1978uw}  
as a natural analogy to the quark sector. 
However, by extension of that analogy, 
the assumption was that this effect, if it existed, 
would be very small.  
As a result, theoretical interest in 1998 was largely
isolated to $CP$ violation. 
In retrospect,  this approach was naive,  but this made sense 
as the guiding principle for the MiniBooNE design at the time.  The goal
was to test a simple two-neutrino oscillation model with equal
probabilities of neutrinos and anti-neutrinos, on the basis that this
would be a good approximation if the underlying reality was BSM
physics.   If a signal was not observed,  the significantly different
systematic errors were expected to result in a clear exclusion of the result.
Thus, the MiniBooNE experiment began running in neutrino mode, 
which provided roughly $\sim$6 times higher rate than anti-neutrino mode--a necessary choice 
since the MiniBooNE experiment was also relied on a significant Booster performance improvement. 
The results showed an anomalous excess of electron like events in the
$\numu$ dominant neutrino mode beam \cite{MB_unexplained} that was similar to,
but not in good agreement with, LSND. 
The experiment then switched to running in anti-neutrino mode,   where
a result in agreement with LSND was observed.    

Rather than consider these events historically,  
we present both results together in the
next section,  followed by a discussion of interpretations and
considerations of follow-up experiments.
There is a world-wide effort to probe the sterile neutrino 
in the region $\De m^2\sim 1~\ueVt$~\cite{sterile}. 
It is desirable for MiniBooNE to confirm this excess is electron-like, 
which is considered the sterile neutrino oscillation signal,  
not background gamma rays associated with $\numu$($\numubar$)NC interactions. 
The MicroBooNE experiment~\cite{uB} was proposed along this line.
The MicroBooNE experiment features a large liquid argon (LAr) time projection chamber (TPC), 
and it has an ability to distinguish an electron (positron) and a gamma ray. 
The MicroBooNE experiment will start data taking in 2014. 
We will discuss more in a later section.

\subsection{The Neutrino and Anti-Neutrino Appearance Oscillation Results}

\begin{figure}[tb]
\begin{center}
\includegraphics[scale=0.5]{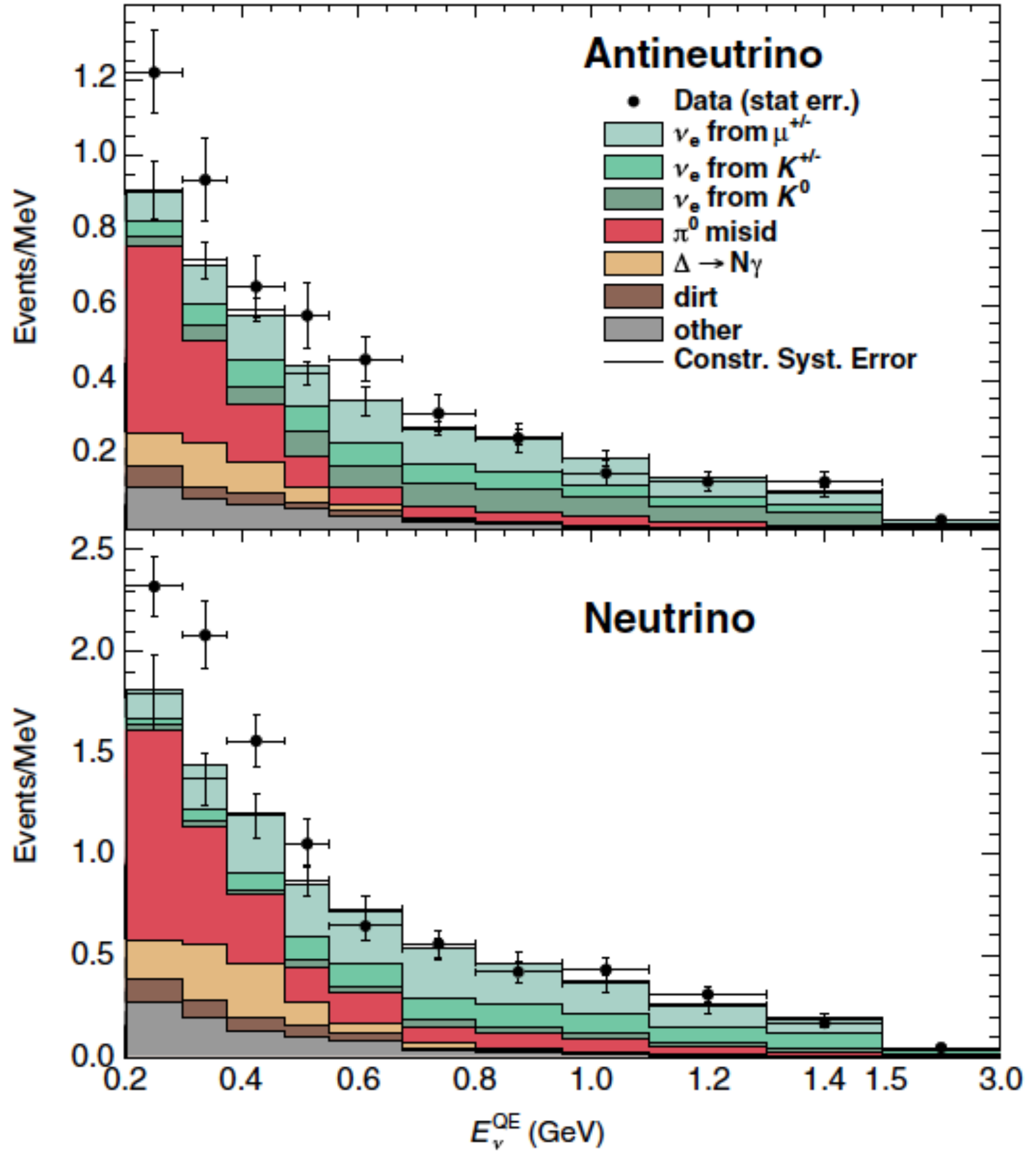}
\includegraphics[scale=0.5]{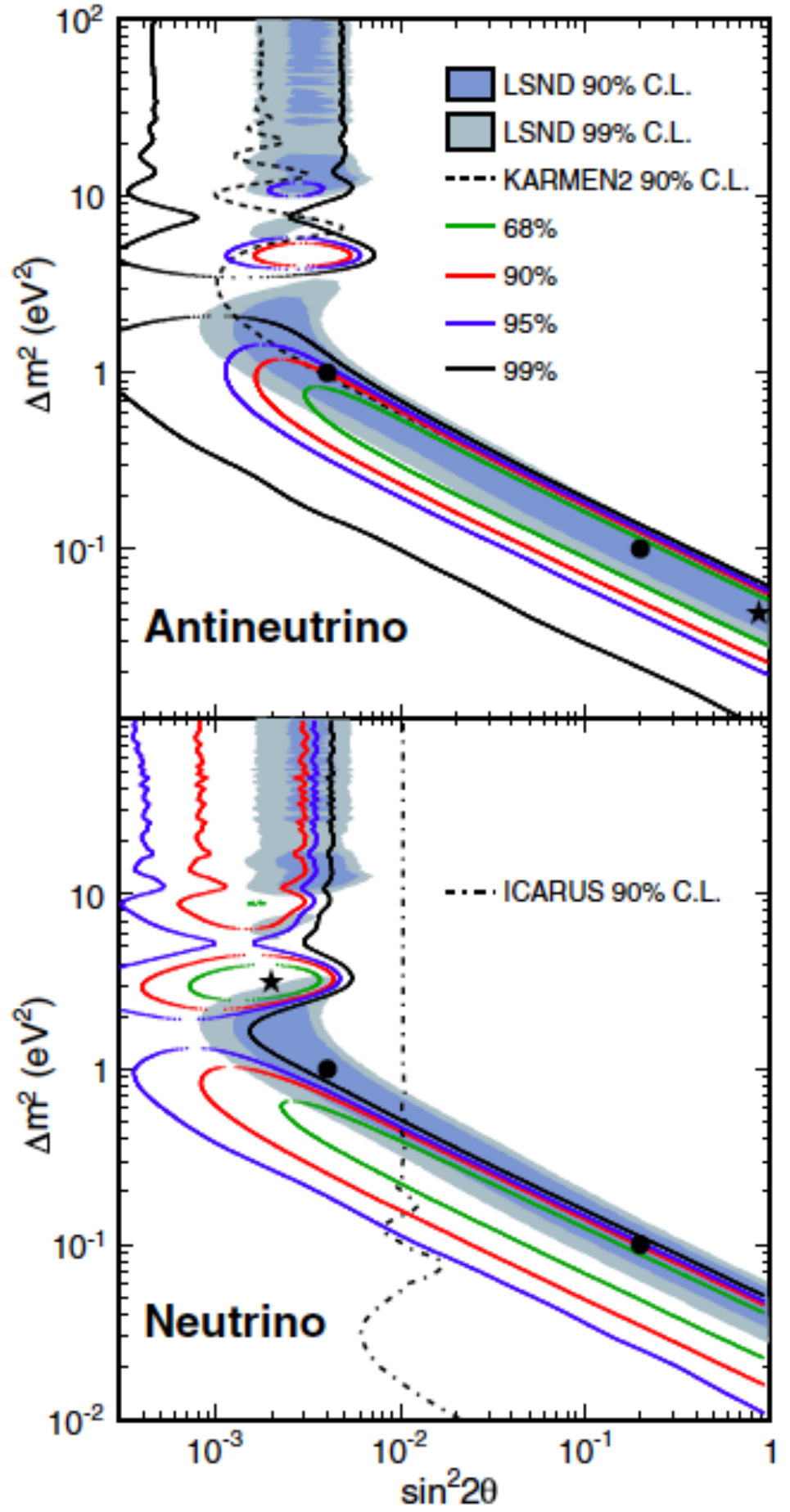}
\end{center}
\caption{\label{fig:osc}
(color online) 
The final MiniBooNE oscillation results~\cite{MB_osc}. 
The left plot shows the reconstructed neutrino energy distribution of oscillation candidate events. 
The top is for anti-neutrino mode and the bottom is for neutrino mode. 
Both modes show excesses in the low energy region, 
while the neutrino mode has higher statistical significance. 
The right plot shows the allowed region in $\De m^2$-$sin^22\th$, 
where the best fit points are shown in black stars. 
The compatibility with the LSND signal is better in anti-neutrino mode.
 }
\end{figure}

After a decade of data collection, MiniBooNE's final appearance
oscillation results have been published~\cite{MB_osc}. 
Figure~\ref{fig:osc} shows the electron candidate ($\numu\to\nue$ oscillation candidate) 
distribution in neutrino mode and positron ($\numubar\to\nuebar$ oscillation candidate) 
distribution in anti-neutrino mode. 
Note that since the MiniBooNE detector is not magnetized, 
in general it cannot distinguish between electrons and positrons, and so both
are grouped into the ``electron-like'' category.

MiniBooNE observed event excesses in both modes of running, 
but the results have slight qualitative difference.
In neutrino mode (left bottom plot), 
there is a statistically significant (3.8$\si$) event excess in the low energy region. 
Although the excess is significant, 
the shape of the spectrum leaves some tension with the oscillation
hypothesis from LSND, 
which you can see from the right bottom plot where the MiniBooNE best fit region 
does not overlap well with the LSND best fit region.    
MiniBooNE uses a  likelihood-ratio technique \cite{MB_osc2010}, 
to find the best fit values ($\De m^2,\sin^2 2\theta)= (3.14~{\rm eV}^2, 0.002)$ 
in neutrino mode, with $\chi^2$/dof of 13.2/6.8.
In anti-neutrino mode (left top plot), 
the observed excess is not as statistically strong as neutrino mode
(2.8$\si$).  This is expected when one compares the protons on target
in each mode and considers the lower anti-neutrino flux and cross
section.   
Although the statistical significance is lower, then shape agreement with
the LSND hypothesis is better. 
Again this can be seen from the right top plot where the parameter space selected by the MiniBooNE data 
agree with the LSND best fit region.   The best fit point in this mode
was (0.05 eV$^2$, 0.842) with $\chi^2$/dof of 4.8/6.9.

The combined result significance is dominated by neutrino mode and is 3.8$\si$.  
It is possible to find compatible regions in a simple
two-neutrino model between the two data sets \cite{MB_osc}.  
However, we emphasize that considering MiniBooNE oscillations 
in the absence of other oscillation experiments lead to misunderstandings. 
We consider this point in a later section.

\subsubsection{Potential Non-oscillation Explanations}

The background-only $\chi^2$-probability for the MiniBooNE oscillation
search was 1.6\% and 0.5\% relative to the
best oscillation fits for neutrino and anti-neutrino mode, respectively. 
Nevertheless, it is important to explore in detail the potential
SM explanations of the MiniBooNE results.  In particular,   
a Cherenkov detector, such as MiniBooNE, lacks the ability to distinguish
electrons from single photons. 
Therefore any single photon production mechanism via neutral current 
interactions is a likely suspect as a background to this search.

The primary source of single photons is the NC1$\piz$ reaction, followed by $\pi^0 \rightarrow \gamma\gamma$, where
one photon is lost because it exits the detector or because the
relativistic boost causes the energy to be too low to allow the
Cherenkov signal to be identified.  At the low energies of MiniBooNE,
the background from two $\piz$ rings that merge is less important than
the case where a photon is lost.    Fortunately,  MiniBooNE has the
largest sample of well reconstructed NC$\piz$ events ever obtained.
Keeping in mind that the largest uncertainties are in the production 
and not in the kinematics of the photons themselves,  MiniBooNE was
able to use this large data set to
carefully evaluate this appearance background~\cite{MB_NCpi0first}. 
This study can constrain the variation of this largest misID background 
(red histogram in Fig.~\ref{fig:osc}, left), 
and we have shown that if NC$\piz$ was the source of the MiniBooNE excess, 
MiniBooNE's systematic error on the production 
would have to be underestimated by an order of magnitude~\cite{MB_unexplained}. 
This is not a likely solution to the problem, and so we turn to single photon production.

MiniBooNE also included the NC single photon process in their simulation. 
The process involves the single photon decay of a neutral current $\De$ resonance, 
which has a small but non-negligible branching ratio ($<$1\% of NC1$\piz$).     
The rate of this process is strongly tied to the resonant production of pions, 
therefore MiniBooNE can utilize their {\it in situ} NC1$\piz$ measurement 
to constrain this background. 
Therefore the variation of this second biggest misID background 
(light brown histograms in Fig.~\ref{fig:osc}, left) 
is also constrained by the NC1$\piz$ measurement, 
and we found this process was not large enough 
to explain the MiniBooNE excess~\cite{MB_unexplained}. 

\begin{figure}[tb]
\begin{center}
\includegraphics[scale=0.3]{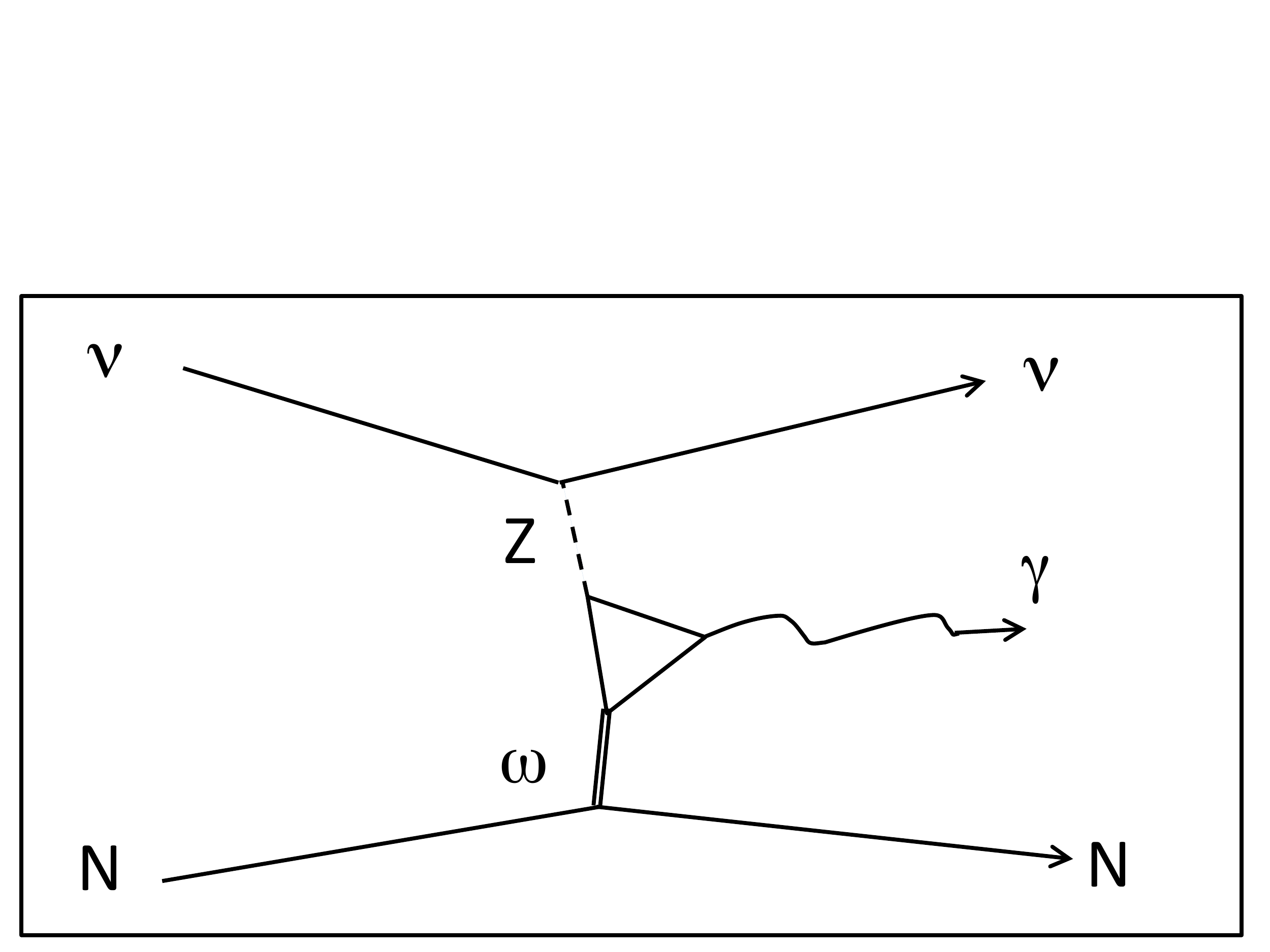}
\end{center}
\caption{\label{fig:anomaly}
The triangular anomaly mediated photon production. 
The neutrino neutral current couples via Z-boson, 
and the target nucleon or nucleus couples with 
a strong force mediated vector meson, such as an omega meson.  
}
\end{figure}

After the first MiniBooNE oscillation result in 2007~\cite{MB_firstosc}, 
it was pointed out that there were additional single-photon-production 
channels missing from the NUANCE \cite{NUANCE} event simulation used
by experiments such as MiniBooNE~\cite{Hill_3H}.  
Figure~\ref{fig:anomaly} shows the relevant underlying diagram. 
This source,  triangular anomaly mediated photon production, 
features weak coupling via the neutrino neutral current, 
and strong coupling with nucleons or nuclei. 
In fact, a similar type of interaction was suggested 
originally in the  1980's~\cite{Khlopov_ncgamma}, 
however, it was not widely noted or further investigated. 
This type of process can generate a single gamma ray from a NC interaction. 
The strength of the anomaly mediated diagram was evaluated~\cite{Hill_ncgamma}, 
and the event rate in MiniBooNE, after convoluting the BNB neutrino flux, 
was, at the time, estimated to be high enough to explain a part of the
MiniBooNE excesses~\cite{Hill_osc}.

The initially high estimate, which may have explained the MiniBooNE 
result, led nuclear theorists to re-evaluate this exotic ``$Z-\ga-\om$
coupling,'' properly including nuclear effects, 
such as Pauli blocking and  $\De$ resonance media width modification, 
as well as including careful calibrations of nuclear parameters from external 
data~\cite{Nieves_ncgamma,Zhang_coh,Zhang_incoh}. 
These are important to include since nuclear effects are sizable in this energy region. 
Note these nuclear effects tend to reduce the cross section. 

\begin{figure}[tb]
\begin{center}
\includegraphics[scale=0.5]{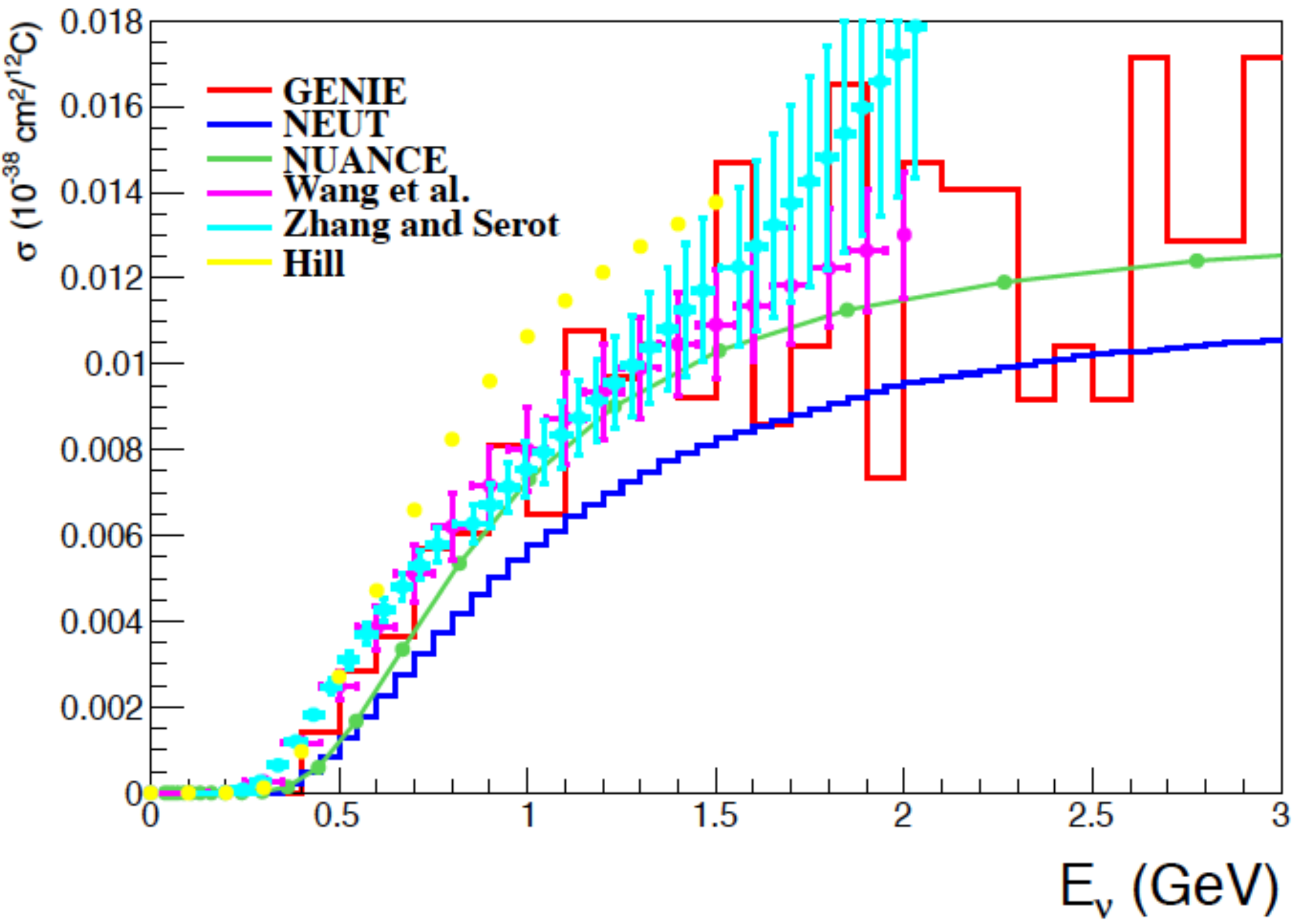}
\end{center}
\caption{\label{fig:MBp_03}
(color online) 
A comparison of the total cross section of NC photon production per $^{12}$C nucleus~\cite{MBp}. 
The neutrino interaction generators used by experimentalists 
(GENIE~\cite{GENIE}, NEUT~\cite{NEUT}, and NUANCE~\cite{NUANCE}) tend to predict lower cross sections 
than state-of-the-art theoretical models 
(Wang {\it et al.}~\cite{Nieves_ncgamma}, 
Zhang and Serot~\cite{Zhang_osc}, 
Hill~\cite{Hill_osc}).
}
\end{figure}

Figure~\ref{fig:MBp_03} shows our current knowledge of this channel~\cite{MBp}. 
The figure shows the total cross section of NC single photon production process per $^{12}$C nucleus, 
which means the cross section includes all potential processes contributing to this final state topology 
(``0 muon + 0 pion + 1 photon + N protons''), both incoherently (neutrino-nucleon interaction) 
and coherently (neutrino-nucleus interaction). 
As you see, all neutrino interaction generators used by experimentalists 
(GENIE~\cite{GENIE}, NEUT~\cite{NEUT}, and NUANCE~\cite{NUANCE}) 
tend to predict lower cross sections 
than state-of-the-art theoretical models by Wang {\it et al.}~\cite{Nieves_ncgamma}, Zhang and Serot~\cite{Zhang_osc}, 
and Hill~\cite{Hill_osc}.

The NC single photon prediction may explain part of
the excess, but it is not likely to explain all of it. 
There was an active discussion on this channel at the recent INT workshop, 
and further experimental data can help to guide more theoretical work~\cite{INT_NCgamma}. 

\begin{figure}[tb]
\begin{center}
\includegraphics[scale=0.7]{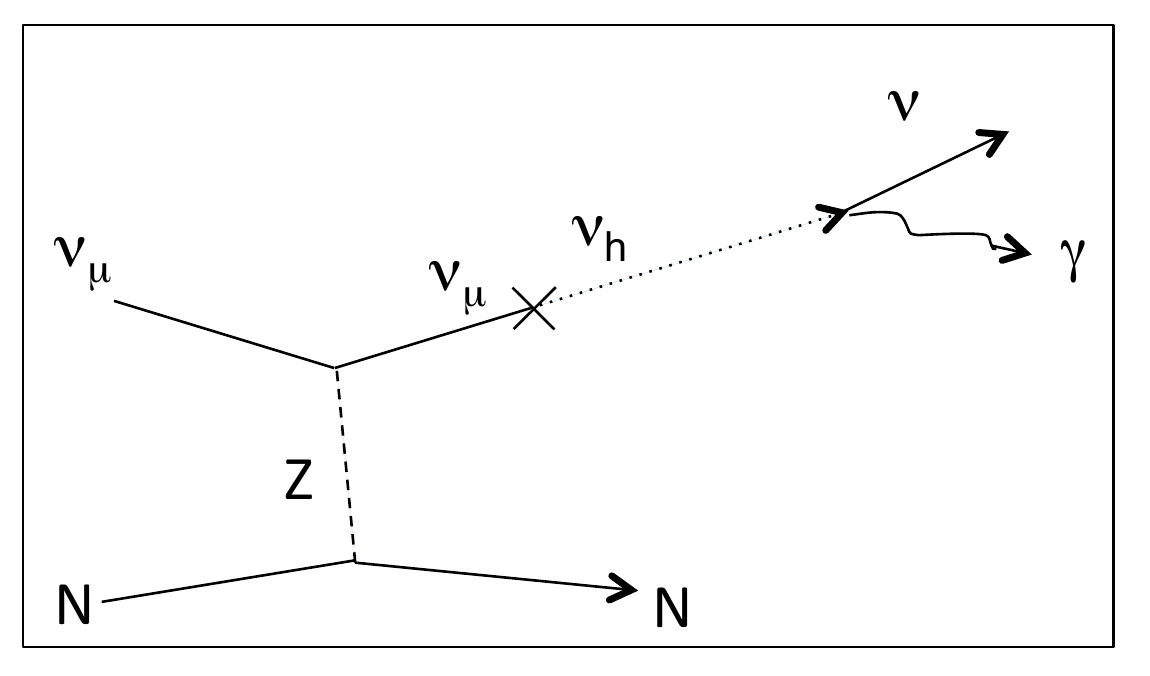}
\end{center}
\caption{\label{fig:MBp_000}
The concept of a heavy neutrino decay signal in the MiniBooNE detector~\cite{Gninenko_osc}. 
The mixing of a neutrino with a hypothetical heavy neutrino and its short life time 
allows for it to decay in the MiniBooNE detector to leave a photon signal.
}
\end{figure}

Meanwhile, a BSM NC single photon model was proposed~\cite{Gninenko_first} 
where a decay of a heavy neutrino produces a single photon signal in the detector. 
Figure~\ref{fig:MBp_000} shows the concept of such a model. 
The heavy neutrino is produced by the mixing with a muon neutrino, 
then the decay of the heavy neutrino leaves a photon signal in the detector. 
Interestingly, the required mass range of the heavy neutrino to produce such a signal in the MiniBooNE detector 
($40~\uMeV<m_h<80~\uMeV$) is not constrained by other experiments. 
The beauty of this model is that  
it also explains the LSND signal, while evading the KARMEN null oscillation result~\cite{Gninenko_osc}. 
 
At this time, NOMAD is the only experiment to have performed 
a dedicated NC single photon search~\cite{NOMAD_NCgamma}. 
The NOMAD result was consistent with their background prediction, thus, NOMAD set a limit on this channel. 
However, the limit was quoted with NOMAD's average energy ($<E>\sim 17~GeV$) 
and is therefore not as relevant for lower energy experiments, such as MiniBooNE. 
Therefore, it is essential for new experiments that seeks to check the MiniBooNE results to
have an ability to distinguish between electrons and photons, such as MicroBooNE~\cite{uB}. 

\subsubsection{Potential Oscillation Explanations}

Numerous articles have been written on the potential of oscillation
models to explain the MiniBooNE signal.  In particular, we recommend
Ref.~\cite{Conrad:2012qt} as a pedagogical discussion of the issues of
fitting the data.   We excerpt the results from this reference here. 

When MiniBooNE and LSND results are considered within the context of
the world's oscillation data,   
$\nu$SM is excluded, because a third mass splitting must be introduced.  
Because the $Z\to\nu \bar \nu$ results from LEP and SLD \cite{PDG}
limit the number of low mass active neutrinos to three,  sterile
neutrinos are introduced to allow for these data sets.    Sterile
neutrinos are a consequence of many theories and can evade limits from
cosmology, as discussed in Ref.~\cite{sterile} .   

If one sterile neutrino is added to the three active neutrinos, then
the model is termed (3+1).    Two additional sterile neutrinos leads
to a (3+2)  model and three results in a (3+3) model.   The mass
states are mixtures of flavor states, and in these models, fits to the
data yield mass states
that are either mostly active flavors or mostly sterile flavors.  The
splitting between the mostly active and mostly sterile flavors is
large, and the splittings between the active flavors is,
comparatively, negligible.   So, in sterile neutrino fits, the
short-baseline approximation where the mostly active flavors are
regarded as degenerate in mass is used.  In such a model, 3+1 models
are simply two-neutrino models, such as what was initially proposed
to explain LSND.

The disagreement between the MiniBooNE neutrino and anti-neutrino 
data leads to very poor fits for 3+1 models.  In order to introduce a
difference in the neutrino oscillation probabilities, $CP$ violation
must be included in the model.  For the term which multiplies the
$CP$-parameter to be significant, there must be two mass splittings
that are within less than two orders of magnitude of each other.
This can be accommodated in a 3+2 model.

Since the MiniBooNE and LSND results were published,  two new
anomalies consistent with high $\De m^2$ oscillations were brought forward.   
These are the reactor anomaly~\cite{Mention_anomaly}, 
which has been interpreted
as $\bar \nu_e \rightarrow \bar \nu_s$, 
and the gallium source anomaly~\cite{Giunti_gallium2} 
which can be interpreted as $\nu_e \rightarrow \nu_s$~\cite{sterile}.  
Both anomalies have weaker significance
than MiniBooNE and LSND, but they can be combined
into a consistent model.

With this said,  many experiments have searched for oscillations in
the high $\De m^2$ region and found no evidence of oscillations.   Ref
\cite{Conrad:2012qt} describes nine such results.     The exclusion
limits for electron-flavor disappearance and electron-flavor
appearance can be shown to be compatible with the results of the four
anomalous measurements.  However, when muon-flavor disappearance is
included,  there is tension between the data sets which leads to low
compatibility, except in the 3+3 picture.

\subsubsection{Near-future Experiment Addressing the MiniBooNE Results}


The MicroBooNE experiment is a large liquid argon time projection chamber (LArTPC) experiment~\cite{uB}. 
It is part of the US LArTPC program~\cite{karagiorgi_LArTPC}, 
with the eventual goal of an ultra-large LArTPC experiment, such as LBNE~\cite{LBNE}. 
The experiments are motivated by the  ``bubble chamber level'' LArTPC imaging quality. 

\begin{figure}[tb]
\begin{center}
\includegraphics[scale=0.5]{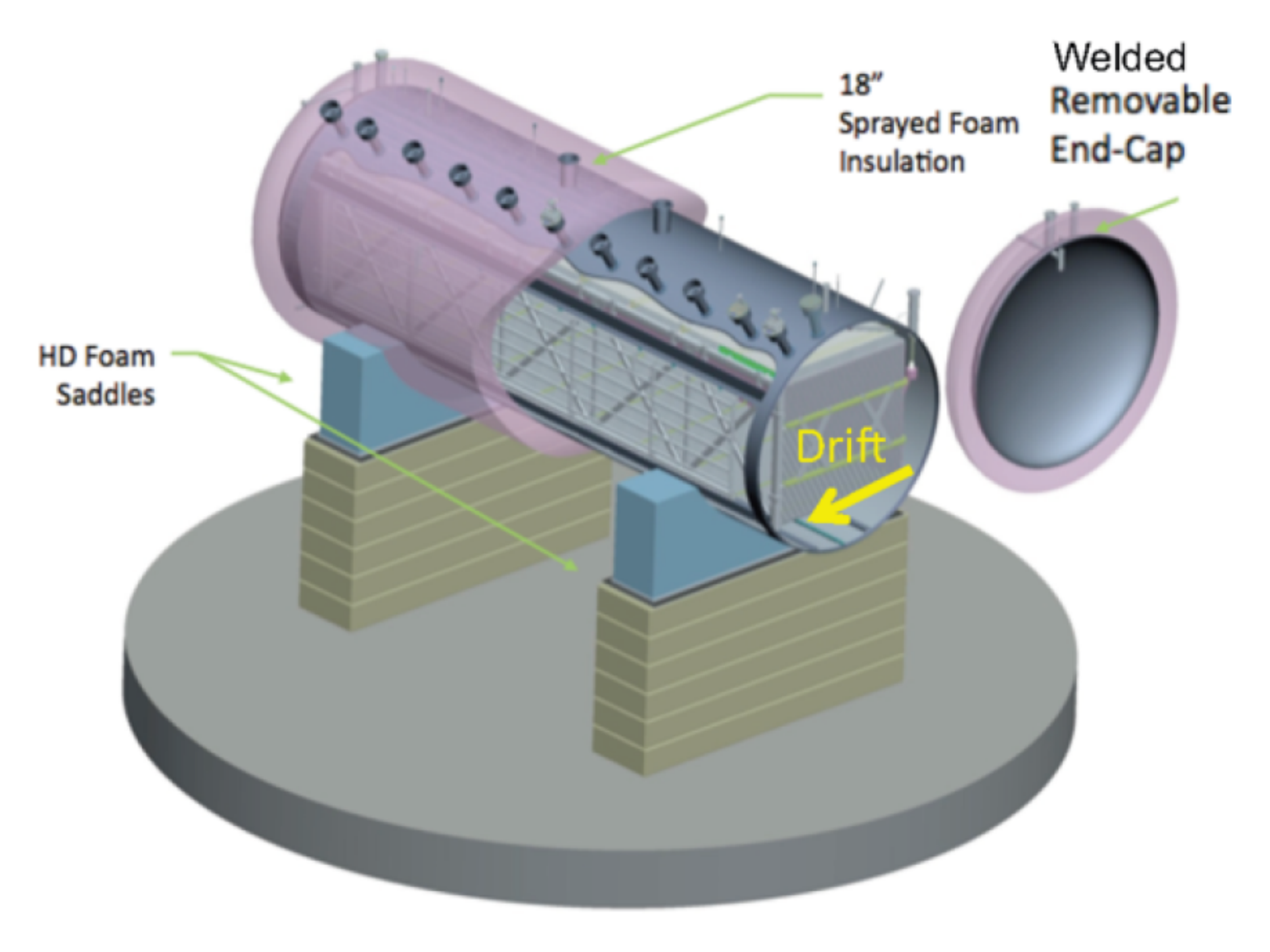}
\end{center}
\caption{\label{fig:uB_detec}
(color online) 
A drawing of MicroBooNE cryostat~\cite{uB}. 
The 170 ton cryostat contains the 89 ton TPC. 
}
\end{figure}

Figure~\ref{fig:uB_detec} shows a drawing~\cite{uB} of 
MicroBooNE's 170 ton foam-insulated cryostat. The TPC volume is 89 tons. 
Ionized electrons along the neutrino-induced charged particle tracks are drifted via a high electric field
in the TPC volume to the anode wires. 
The node wires are configured on three planes alternating by $60^{\circ}$ orientation, 
to allow 3-dimensional reconstruction of the tracks. 
The first 2 wire planes record the signal from the induction on wires, 
and the last plane records the actual collection of ionization electrons. 

An array of 8-inch PMTs is equipped behind the wire planes~\cite{uB_PMTsystem}. 
The main purpose of this photon collection system is to reject
out-of-time cosmic rays and to trigger on in-time signals, 
since the scintillation light from the interaction arrives in $\sim$ns whereas 
the time scale of ionization electron drift is of order $\sim$ms. 
The detection of scintillation photons from LAr is not straightforward. 
First of all, the wavelength of Ar scintillation light is 128~nm,  
which requires careful R\&D on potential wavelength shifters for use in LAr~\cite{Tad,Chiu,Ben_benzophenone}. 
Second, the PMTs themselves behave differently in a cryogenic
environment as compared to a warm environment, 
leading to the need for careful characterization \cite{uB_PMTtest}.

The purity of the liquid argon must be kept very high to allow electrons to drift a long distance. 
Electro-negative impurities (e.g. water and oxygen molecules) are removed through a custom made filter 
to achieve $\leq$~ppb level impurity~\cite{Pordes_filter,Pordes_luke}. 
Such filtering is also effective for removing nitrogen molecules, 
which don't affect electron drift but do attenuate scintillation light~\cite{Ben_nitrogen}.

A high resolution LArTPC detector will be a powerful tool in understanding the MiniBooNE signal,  
because the detector is expected to have the excellent electron-photon separation.
Energetic electrons and photons both produce an electromagnetic shower in a
LArTPC. However, the initial $\frac{dE}{dx}$ of a single photon will be twice higher 
than in the single electron case in the first few centimeters before the
track develops into the shower. Due to their  
high resolution capabilities, LArTPC detectors can distinguish this difference. 
Moreover, a displaced vertex, in the case of a photon conversion, 
can be distinguished from a track that is continuous from the vertex, indicative of an electron. 
The combination of these details can provide high efficiency background rejection for MicroBooNE.
 
\section{Test of Lorentz and CPT violation}

Lorentz and CPT violation are scenarios motivated from Planck scale theories, 
such as string theory~\cite{SLSB}. 
In the effective field theory framework, 
Lorentz violation contributes additional terms to the vacuum Lagrangian of neutrinos, 
and hence modifies neutrino oscillations~\cite{Mewes_1,Diaz_LBA}.  
Since Lorentz violating fields are of fixed-direction in the universe, 
if Lorentz invariance is broken, the rotation of the Earth causes 
a sidereal time dependence of neutrino oscillation signals.
There are number of phenomenological neutrino oscillation models 
based on Lorentz and CPT violation~\cite{Mewes_2,Diaz_puma1,Diaz_puma2}, 
some of which can explain the LSND excess~\cite{tandem}. 
In fact, a sidereal time dependence analysis of LSND data~\cite{LSND_LV} failed 
to reject the Lorentz violation scenario. 
Therefore, it might be possible to reconcile LSND and MiniBooNE oscillation signals 
under Lorentz violation. 

\subsection{Analysis}

Although Lorentz violation can be studied in any frame or coordinate system, 
it is convenient to choose one coordinate system to compare data sets. 
The standard choice is the Sun-centered celestial equatorial coordinates~\cite{table_LV}, 
where the origin of the coordinate is the center of the Sun.  
The orbital plane of the Earth is tilted so that the orbital axis and the rotation axis 
of the Earth align. This direction define the Z-axis. 
The X-axis points vernal equinox, and the Y-axis is chosen to complete the right handed system. 
Because the time scale of the rotation of the galaxy is too long for any terrestrial experiments,  
the Sun-centered frame is the better choice to test rotation symmetry 
(by using the rotation of the Earth) and Lorentz boost (by using the revolution of the Earth). 

Having defined the coordinates, one uses the Standard Model-Extension (SME)~\cite{SME_1,SME_2,SME_3} 
as the framework for a general search for Lorentz violation. 
The SME can be considered a minimum extension of the SM, including the particle Lorentz and CPT violation. 
For the neutrino sector, the SME Lagrangian can be written as~\cite{Mewes_1},
 
\beq
{\cal L} 
&=& 
\fr{1}{2}i{\bar {\ps}}_A{\Ga}^{\mu}_{AB}\stackrel{\leftrightarrow}
{D_{\mu}}{\ps}_{B}-{\bar {\ps}}_{A} M_{AB}{\ps}_{B}+h.c.,\\
\Ga_{AB}^{\nu}&=&
\ganu\de_{AB}+c_{AB}^{\mu\nu}\gam+d_{AB}^{\mu\nu}\gaf\gam+e_{AB}^{\nu}+if_{AB}^{\nu}\gaf
+\fr{1}{2}g_{AB}^{\la\mu\nu}\si_{\la\mu},~\label{eq:gamma}\\
M_{AB}&=&
m_{AB}+im_{5AB}\gaf+a_{AB}^{\mu}\gam+b_{AB}^{\mu}
+\fr{1}{2}H_{AB}^{\mu\nu}\si_{\mu\nu}.~\label{eq:mass}
\eeq

Here, the $AB$ subscripts represent the flavor basis. 
The first term of Eq.~\ref{eq:gamma} and the first and second terms of
Eq.~\ref{eq:mass} are the only nonzero terms in the SM, 
and the rest of the terms are from Lorentz violation. 

The physics consequences predicted by Lorentz violation are very rich. Among them, 
we are interested in Lorentz violating neutrino oscillations, 
where neutrino oscillations act natural interferometers 
sensitive to small space-time physics. 
The smoking gun of Lorentz violation is 
the sidereal time dependence of physics observables. 
Therefore, we used the Lorentz violating neutrino oscillation formula~\cite{Mewes_3}
to fit the sidereal time distribution of the MiniBooNE oscillation candidate data.  
Here, potentially, any day-night effect, either from the beam or from the detector, 
could mimic the sidereal time distribution. 
MiniBooNE studied effects versus the time distribution of the delivered POT 
and the high statistics $\numu$($\numubar$) CCQE sample~\cite{MB_CCQE,MB_ANTICCQE}, 
and confirmed that day-night effects on both $\nue$ and $\nuebar$ 
oscillation candidates are well below statistical errors.

\begin{figure}[tb]
\begin{center}
\includegraphics[scale=0.6]{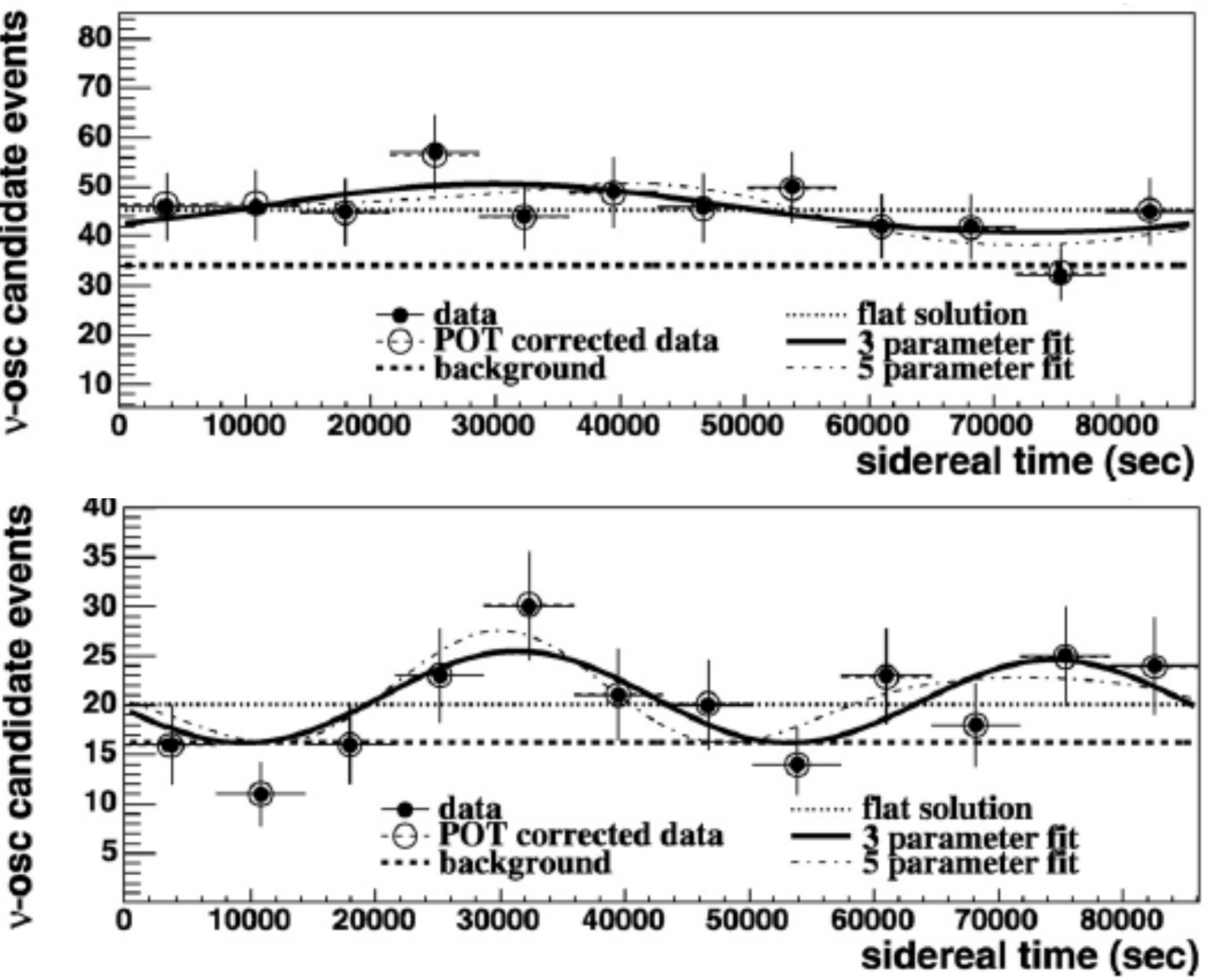}
\end{center}
\caption{\label{fig:LV}
The MiniBooNE Lorentz violation results~\cite{MB_LV}. 
The top plot shows the neutrino mode electron-like 
low energy excess sidereal time distribution, 
and the bottom plot shows the anti-neutrino mode sidereal time distribution. 
Here, the data with a POT correction (open circle) show the size of the beam day-night variation. 
There are three fit curves based on different assumptions, a flat solution (dotted), a
three parameter fit (solid curve), and a full five parameter fit (dash-dotted curve).
}
\end{figure}

\subsection{Results}

Figure~\ref{fig:LV} shows the neutrino and anti-neutrino mode electron-like events 
as a function of sidereal time~\cite{MB_LV}. 
There are three curves fit to the data, 
depending on different hypotheses. 
Although the anti-neutrino mode electron-like events show a rather interesting sidereal time dependence, 
the statistical significance is still low. 
Therefore, MiniBooNE found that the data are consistent with no Lorentz violation. 
This analysis provided the first limits on five time independent SME coefficients, 
at the level of $10^{-20}$~GeV (CPT-odd) and order $10^{-20}$ (CPT-even). 
Further analysis inferred limits on each SME coefficient, 
and, together with limits from the MINOS near detector~\cite{MINOS_LV_ND,MINOS_LV_NDanti}, 
it turns out these limits leave tension to reconcile the MiniBooNE 
and LSND data sets under a simple Lorentz violation motivated scenario~\cite{Teppei_LV}. 

In fact, existing limits from MiniBooNE~\cite{MB_LV}, 
MINOS~\cite{MINOS_LV_ND,MINOS_LV_NDanti,MINOS_LV_FD,MINOS_LV_FDanti}, IceCube~\cite{IceCube_LV}, 
and Double Chooz~\cite{DC_LV,DC_LVanti} set very tight limits 
on possible Lorentz violation in the neutrino sector at the terrestrial level. 
This was one of the reasons why the superluminal neutrino 
signal from OPERA~\cite{OPERA_LV} was suspicious from the beginning.
Such a signal would have required very large Lorentz violation,
while avoiding all these constraints when writing down the theory.
Strictly speaking, this is possible--limits on Lorentz violation from the oscillation experiments 
cannot be applied directly to the neutrino time of flight (TOF) measurements~\cite{Mewes_HighOrderNu}. 
However, introducing large Lorentz violation in the neutrino TOF
without other large parameters such as those associated with
oscillations seems unnatural.

\section{Dark Matter Search \label{DMsearch}}

The proton collisions on target in the BNB line that produce 
a large flux of neutrinos could, potentially, 
produce sub-GeV scale dark matter particles, 
that mimic NCE interactions in the MiniBooNE detector~\cite{Batell_1,deNiverville_1,deNiverville_2}.
The most interesting scenario is that this light dark matter particle is the dark matter of the universe, 
which requires a light vector mediator particle (called a ``dark photon''), 
in the model in order to obtain an efficient annihilation cross section. 
The minimum extension of the SM with the light dark matter particle  
and the vector mediator can be written in the following way~\cite{deNiverville_1}, 
\beq
{\cal L} 
&=& {\cal L}_{SM}-\frac{1}{4}V_{\mu\nu}^2+\frac{1}{2}m_V^2V_{\mu}^2+
\ka V_{\nu}\partial_\mu F^{\mu\nu}+|D_{\mu}\ch|^2-m_\ch^2|\ch|^2+\cdots~.
\eeq
The model has four free parameters:
the mass of the light dark matter $m_\ch$, 
the mass of the vector mediator $m_V$, 
kinetic mixing of the vector mediator and the photon $\ka$, 
and the vector mediator's gauge coupling $e'$ (or $\al'=\frac{e'^2}{4\pi}$). 
Nonzero $\ka$ leads to the decay of neutral mesons to a photon and a dark photon, 
and the dark photon in turn can decay to dark matter particles. 
This would be the dominant process to produce dark matter particles in the BNB. 
The second process is direct production from the parton level 
annihilation by protons colliding in the target. 

\subsection{MiniBooNE Searches for Dark Matter particles}

\begin{figure}[tb]
\begin{center}
\includegraphics[scale=0.4]{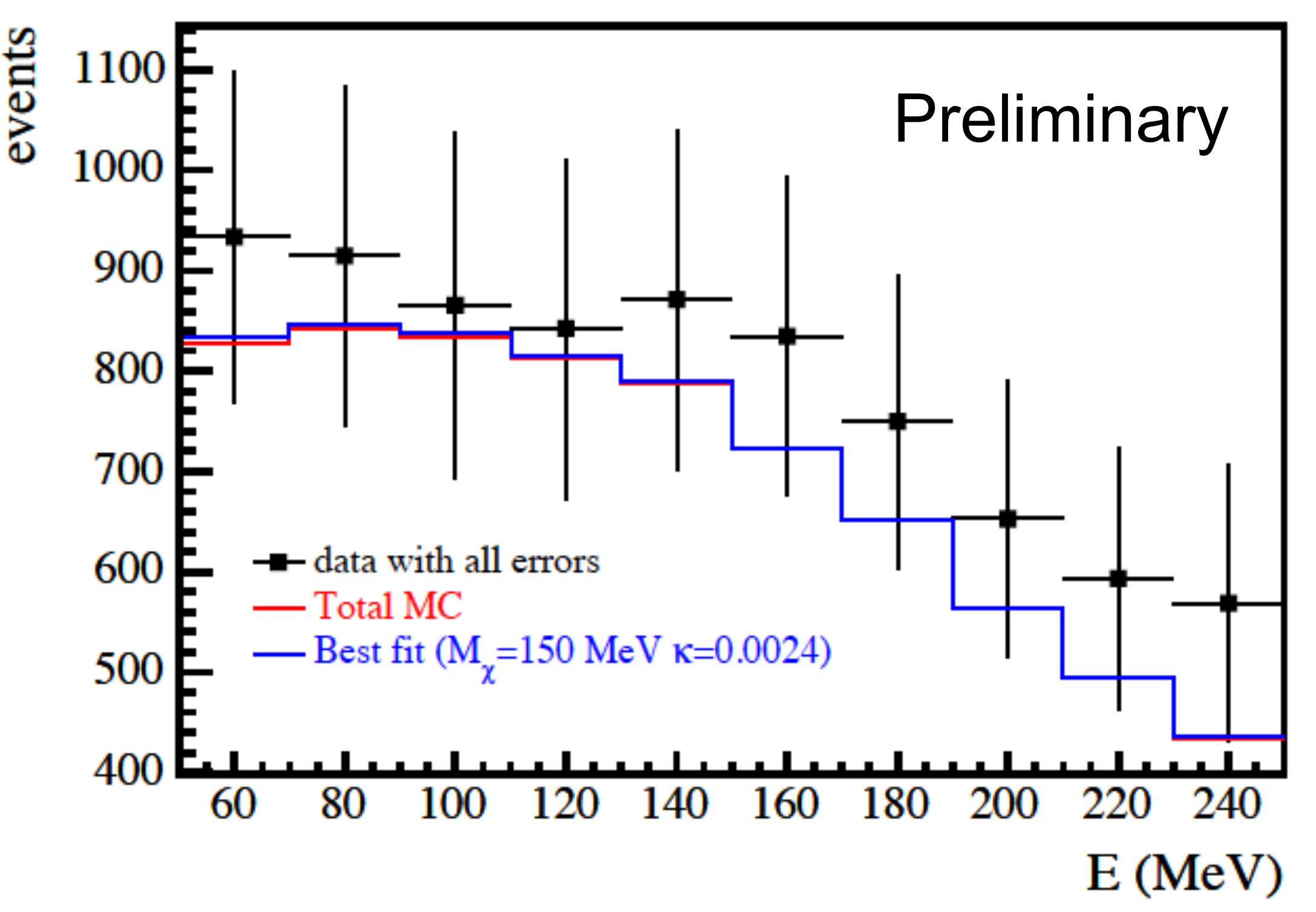}
\end{center}
\caption{\label{fig:dm_15}
(color online) 
The dark matter fit result to the NCE data~\cite{MBd}. 
The plot shows the total energy distribution of the anti-neutrino NCE sample, 
with a light dark matter hypothesis fit (best fit values, $M_{\ch}=150~\uMeV$ and $\ka=0.0024$).
As can be seen, the current sensitivity to the light dark matter model is low. 
}
\end{figure}

MiniBooNE tested this model with the existing anti-neutrino NCE data
set, taken during the oscillation studies.    
Figure~\ref{fig:dm_15} shows the fit result 
with a light dark matter particles hypothesis~\cite{MBd}. 
The anti-neutrino mode data
set is used because it has a lower neutrino interaction rate than the neutrino mode beam.
Nevertheless, 
due to the anti-neutrino backgrounds, 
only weak limits are obtained on the kinetic mixing parameter $\ka$. 

\begin{figure}[tb]
\begin{center}
\includegraphics[scale=0.5]{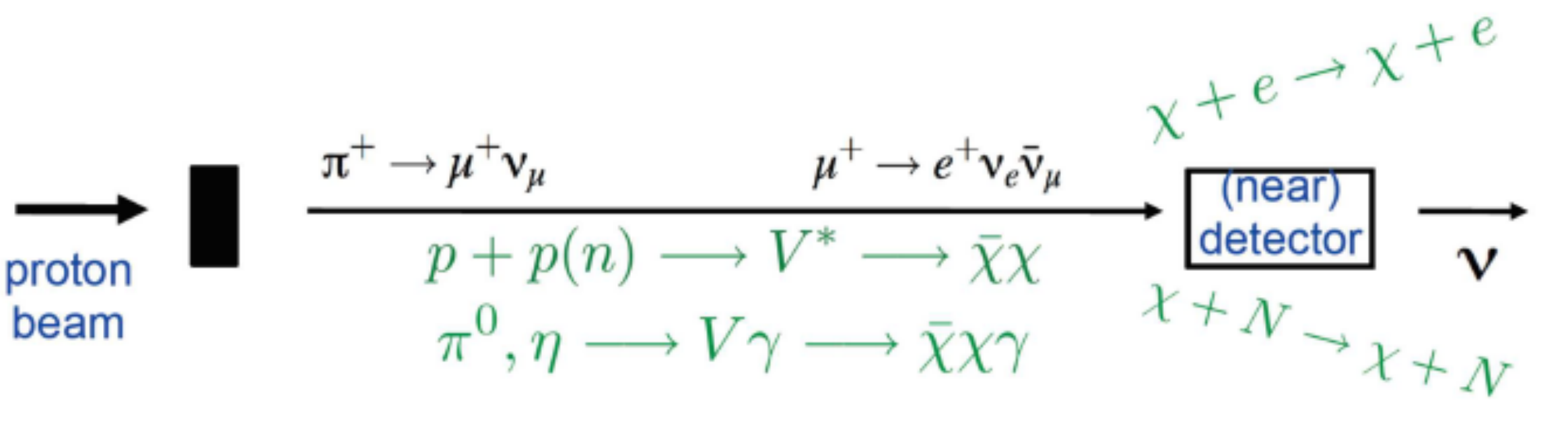}
\end{center}
\caption{\label{fig:dm}
(color online) 
The concept of the dark matter beam in MiniBooNE~\cite{MBd}. 
The dominant production mode of dark matter particles are decays of the mediator particles 
created by decays of neutral mesons. 
The dark matter particles can be also made through 
the direct collisions of protons on the beam dump.
}
\end{figure}

This motivated a tuning of the proton beam that 
allowed MiniBooNE to run in a mode in which the protons are 
directed on to the beam dump instead of the target, 
eliminating the DIF neutrino flux. 
Figure~\ref{fig:dm} shows the schematic of this measurement~\cite{MBd}. 
The beam-dump mode is achieved by tuning 
the $\sim$1~mm beam to aim 0.9 cm gap between 
the beryllium target rod and the inner conductor of the horn, 
to hit the beam dump located at the end of decay pipe 
(50~m from the target) directly. 
This reduces the neutrino
background by roughly a factor of 67. 
Dark matter production is largely unaffected in this run mode since 
it occurs through neutral meson decay. 
MiniBooNE is now running in this configuration.
The goal is to accumulate $1.75\times 10^{20}$~POT data before MicroBooNE starts beam data taking 
in the neutrino mode, not the beam-dump mode. 

\subsection{Parameter space of light dark matter particles and vector mediators}

\begin{figure}[tb]
\begin{center}
\includegraphics[scale=0.6]{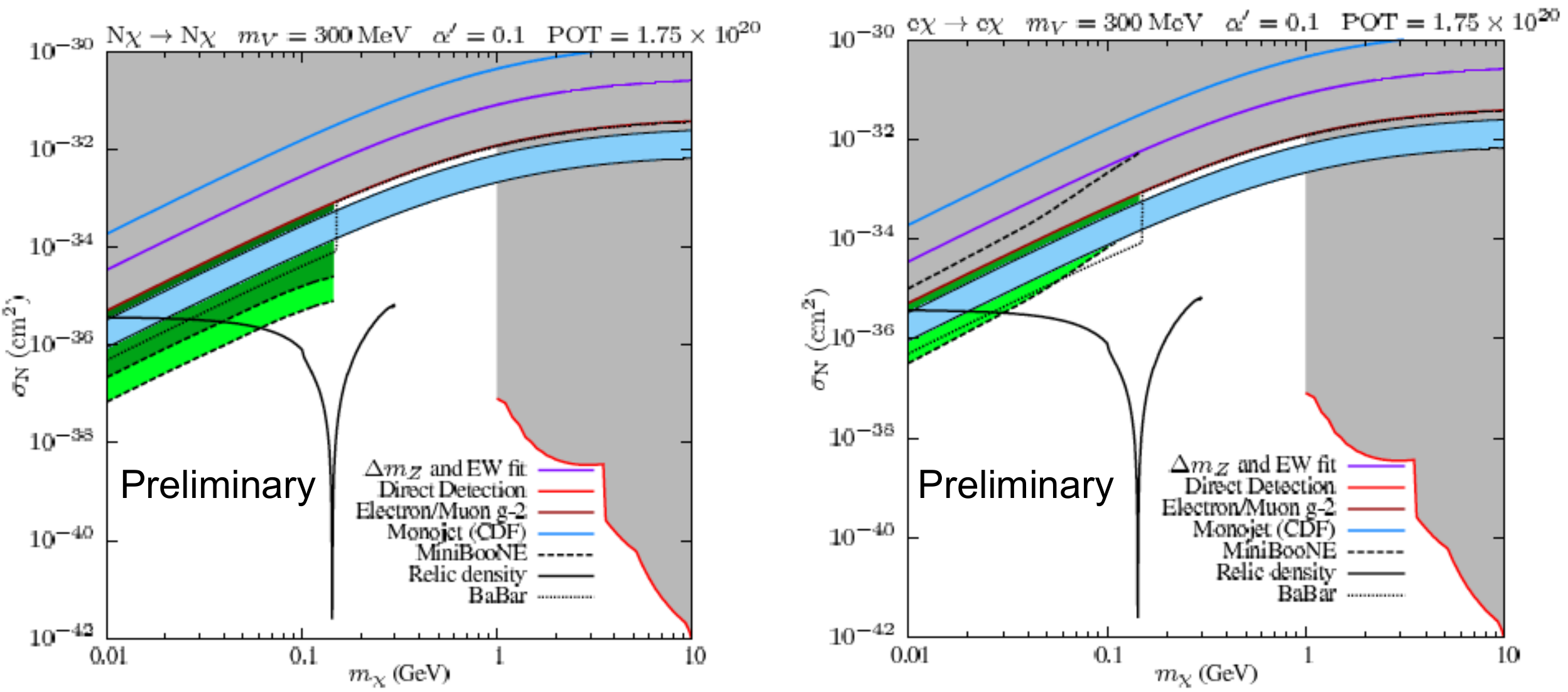}
\end{center}
\caption{\label{fig:dm_04}
(color online) 
The MiniBooNE dark matter particles search phase space~\cite{MBd}. 
Here, the x-axis is the dark matter mass $m_\ch$, 
and the y-axis is either the dark matter-nucleon or dark matter-electron cross section, 
assuming the vector mediator mass and the gauge coupling 
($m_V=300~\uMeV$ and $\al’=0.1$). 
The MiniBooNE exclusion region can be seen in green.
}
\end{figure}

\begin{figure}[tb]
\begin{center}
\includegraphics[scale=0.6]{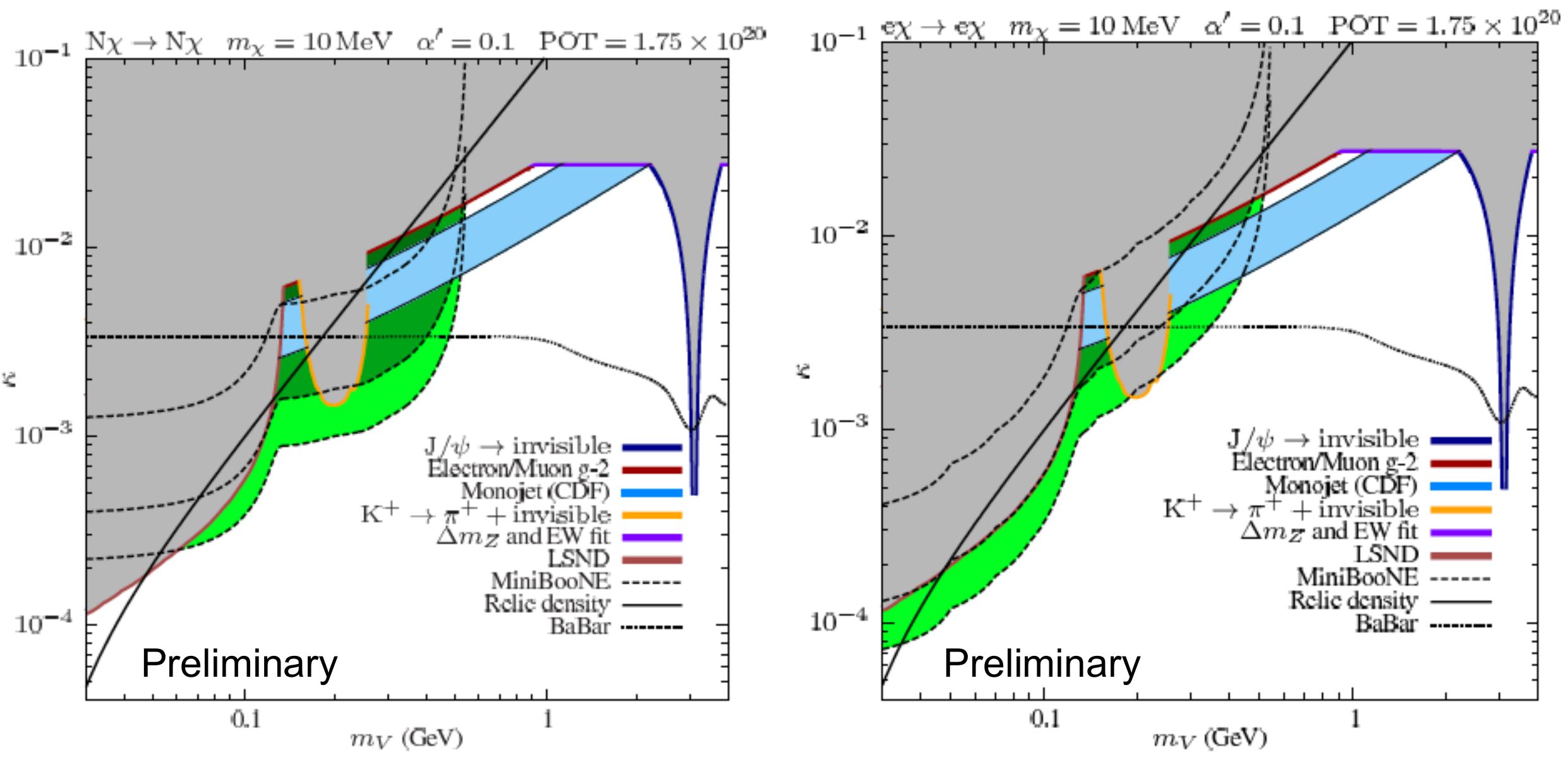}
\end{center}
\caption{\label{fig:dm_06}
(color online) 
The MiniBooNE dark matter search phase space~\cite{MBd}. 
Here, x-axis is the vector mediator mass $m_V$, 
and the y-axis is the kinetic mixing parameter $\ka$, 
assuming the dark matter mass and the gauge coupling 
($m_\ch=10~\uMeV$ and $\al’=0.1$).  
MiniBooNE exclusion region can be seen in green.
}
\end{figure}

Figure~\ref{fig:dm_04} shows the two-dimensional phase space of dark matter-nucleon and 
dark matter-electron scattering cross sections vs. dark matter mass $m_\ch$~\cite{MBd}. 
The limits from direct searches end up at the right side ($m_\ch\sim$1~GeV), 
and the left-side light dark matter region is explored by other techniques, 
such as rare decays and collider physics. 
MiniBooNE addresses direct light dark matter searches. 
In the case of either interaction, 
MiniBooNE is sensitive to the dark matter mass in the 10 to 200 MeV mass region.

There are many reasons why such a light dark matter search is interesting. 
First, recent data~\cite{DAMA,CoGeNT,CRESSTII,CDMSII_Si} 
from the direct WIMP (weakly interacting massive particle) searches suggest 
possible signals of dark matter particles in the lighter mass region.  
For example, SuperCDMS is also aiming the low mass dark matter search 
by utilizing the ionization signals~\cite{SuperCDMS}.  
Second, the muon g-2 anomaly can be explained by the presence of a vector mediator~\cite{g-2,Pospelov_g-2}. 
Although the interesting phase space of muon g-2 was already excluded by other experiments, 
MiniBooNE can further push the limits in this region. 

The sensitivity that is obtained from the dark matter-electron 
scattering looks weaker than dark matter-nucleon 
in the $\si-m_{\ch}$ phase space (Fig.~\ref{fig:dm_04}, right), 
however, as Figure~\ref{fig:dm_06} shows, 
the limit from the dark matter-electron interaction 
can be stronger in the low vector mass region in $\ka-m_V$ phase space~\cite{MBd}. 
Therefore, both channels are complimentary and MiniBooNE should strive to measure both. 
There was a little interest in $\nu$-e elastic scattering because of its small cross section, 
but this electron channel is as important as the nucleon channel for the dark matter search.

\begin{figure}[tb]
\begin{center}
\includegraphics[scale=0.5]{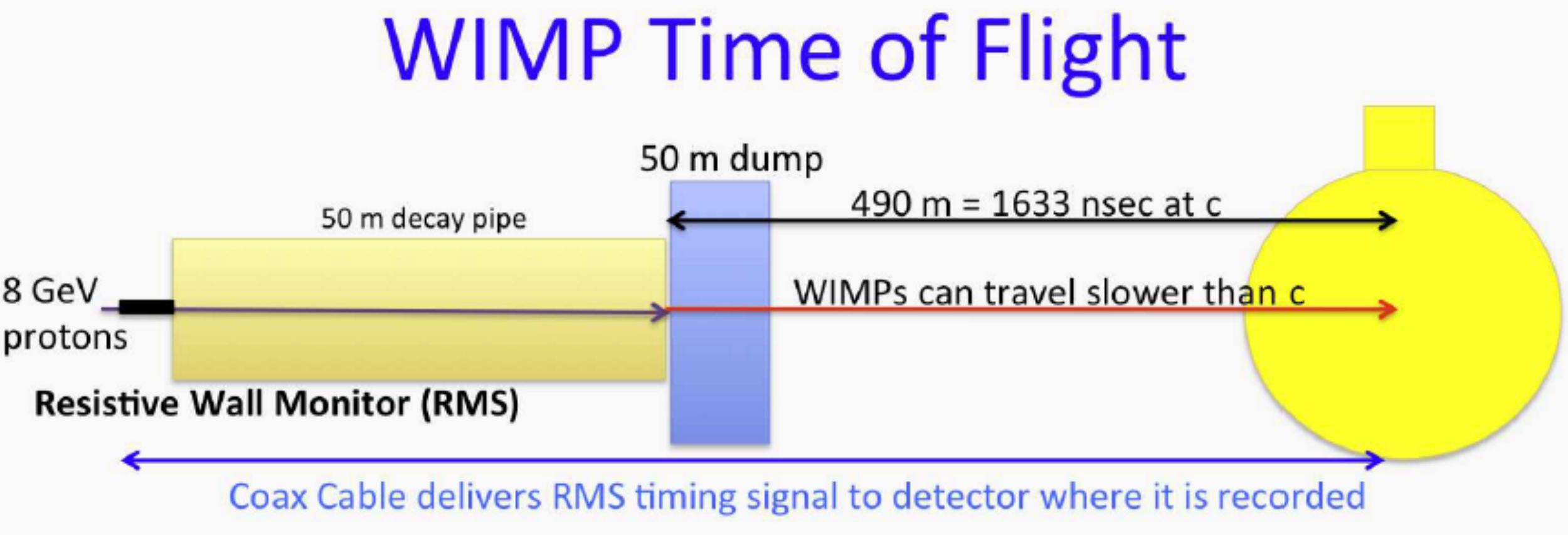}
\end{center}
\caption{\label{fig:dm_08}
(color online) 
The concept of dark matter particles TOF.
Because of the localization of the dark matter particle production in time and in space, 
the dark matter beam has well-defined timing structure. 
}
\end{figure}

\subsection{Dark matter time of flight (TOF)}

MiniBooNE's sensitivity to dark matter particles can be further improved by
combining event topology and kinematics with the timing information.
Figure~\ref{fig:dm_08} shows the ``dark matter TOF'' concept. 
The dark matter particles are most likely produced at the beam dump after 
prompt decays of neutral pions or etas ($<10^{-16}$~sec), 
so the dark matter production is localized in both time and space. 
This would result in a dark matter beam that has a well-defined timing, 
and allows us to perform the TOF-based searches. 
The heavier dark matter particles should be slower than the neutrinos 
(as well as the speed of light).   
Thus the dark matter particles would lag behind the bunch center and separate
from the neutrino background.

\begin{figure}[tb]
\begin{center}
\includegraphics[scale=0.4]{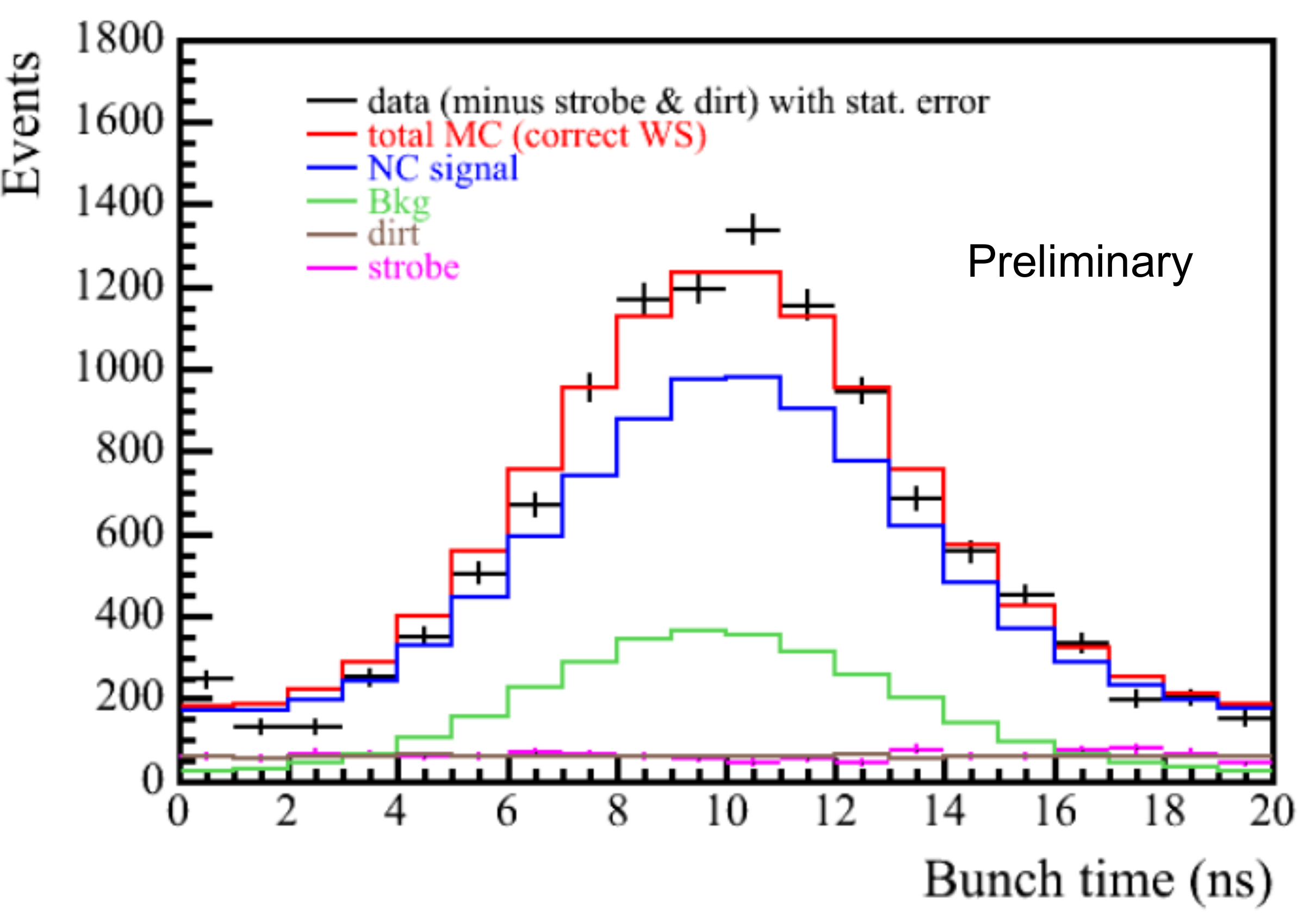}
\end{center}
\caption{\label{fig:dm_10}
(color online) 
The reconstructed NCE event time profile for the anti-neutrino mode beam. 
The events are overlaid relative to the bunch center. 
As expected, the data peaks in the bunch center, which means these are dominated with 
anti-neutrino interactions and there is no delay of events. 
}
\end{figure}

In the Fermilab Booster, the 81 bunches have 19~ns separations (Sec~\ref{sec:beam}).
MiniBooNE defines events within 4~ns$<T<16$~ns from the bunch center as the 
in-time events, and the $T<4$~ns and $T>16$~ns events are out-time. 
The absolute timing information of all bunches is recorded by 
the resistive wall monitor (RWM) which is located just before the
target. 
Using the previous MiniBooNE anti-neutrino run to test this idea, 
Figure~\ref{fig:dm_10} shows the overlaid profile of all bunches of anti-neutrino NCE candidate events~\cite{MBd}. 
As expected, the data shows the peak in in-time, 
because the data is dominated by anti-neutrino NCE interactions.   

A beam dump test run was performed for one week during 2012 running. 
During the beam-dump mode test run, 
the timing of neutrinos was tested using CC interaction. 
Since the CC interaction is detected through the prompt Cherenkov light from the muons, 
timing resolution is better than NCE events. 
Using the new system installed for the beam-dump run, MiniBooNE achieved 1.5~ns resolution~\cite{MBd}. 
The resolution will be worse for NCE because of the nature of the exponential decay of scintillation light, 
but MiniBooNE, nevertheless, still expects $\sim$4~ns resolutions. 
This gives full confidence for MiniBooNE to perform a full beam-dump run.

\section{Conclusion}

Since beginning its run in 2002, MiniBooNE has been searching for new
physics in a wide variety of ways.   The most important results have
been those related to oscillations of sterile neutrinos, which has
pushed the community toward new and exciting experiments 
in the future \cite{uB,IsoDAR,NuSTORM,Cao_reactor,sterile}.  
MiniBooNE also tested for possible signals from the Planck scale, {\it i.e.} Lorentz violation, 
and set very strong constraints on the SME coefficients.  
MiniBooNE's light dark matter search with a beam-dump configuration run is a
unique opportunity that can provide the best limit on the dark matter
mass in the 10 to 200 MeV range.  All of these searches have been
grounded in the revolutionary set of cross section measurements
performed with MiniBooNE.  This experiment demonstrates the 
rich possibilities to go beyond the standard model in low cost
short-baseline venues and encourages a strong investment in future programs.

\begin{center}
{ {\bf Acknowledgements}}
\end{center}
JC thanks the National Science Foundation for support through NSF-PHY-1205175.
We thank Brian Batell for inputs about light dark matter physics, also 
we thank Joshua Spitz for careful reading of the manuscript and valuable comments.

  \bibliography{MBreview}
  \bibliographystyle{kickass}


\end{document}